\documentclass{IEEEtran}
\usepackage{amsmath,amssymb,amsfonts}
\usepackage{algorithmic}
\usepackage{graphicx}
\usepackage{textcomp}
\def\BibTeX{{\rm B\kern-.05em{\sc i\kern-.025em b}\kern-.08em
    T\kern-.1667em\lower.7ex\hbox{E}\kern-.125emX}}

\usepackage[framed]{matlab-prettifier}

\usepackage{xcolor}

\usepackage{tikz}
\usepackage{circuitikz}
\usetikzlibrary{math}
\usepackage{mathrsfs}

\usepackage{siunitx}

\usepackage{placeins}

\ifCLASSOPTIONcompsoc
 \usepackage[caption=false,font=normalsize,labelfont=sf,textfont=sf]{subfig}
\else
 \usepackage[caption=false,font=footnotesize]{subfig}
\fi

\usepackage{pgfplots}
\pgfplotsset{compat=1.18} 
\usepgfplotslibrary{colorbrewer}
\usetikzlibrary{pgfplots.colorbrewer}
\usetikzlibrary{plotmarks}
\usetikzlibrary{arrows.meta}
\usetikzlibrary{spy}
\usetikzlibrary{patterns}
\usetikzlibrary{patterns.meta}
\usetikzlibrary{fit, backgrounds}
\usetikzlibrary{positioning}

\usepackage{cite}

\newcommand{\figref}[1]{Fig.~\ref{#1}}
\newcommand{\secref}[1]{Section~\ref{#1}}

\newcommand{\id}[0]{\mathbb{I}}

\newcommand{\mli}[2]{#1^{[#2]}}
\newcommand{\e}[0]{\revI{\text{e}}}

\newcommand{\inv}[1]{#1^{-1}}
\newcommand{\rev}[1]{#1}
\newcommand{\revI}[1]{#1}
\newcommand{\revII}[1]{#1}
\newcommand{\revIII}[1]{#1}
\newcommand{\revIV}[1]{#1}
\newcommand{\revV}[1]{#1}
\newcommand{\portI}[1]{I_{p,#1}}
\newcommand{\portV}[1]{V_{p,#1}}
\newcommand{\multX}[0]{\mathord{\times}}
\newcommand{\ju}{\revI{\mathrm{j}\mkern1mu}}
\newcommand{\Zp}[1]{\revI{Z_{p,#1}}}
\newcommand{\Za}[1]{\revI{Z_{a,#1}}}
\newcommand{\colS}[1]{\revI{\mathcal{C}_{#1}}}
\newcommand{\elDist}[0]{\Delta_n}

\pgfplotsset{
  log x ticks with fixed point/.style={
      xticklabel={
        \pgfkeys{/pgf/fpu=true}
        \pgfmathparse{exp(\tick)}%
        \pgfmathprintnumber[fixed relative, precision=3]{\pgfmathresult}
        \pgfkeys{/pgf/fpu=false}
      }
  },
  log y ticks with fixed point/.style={
      yticklabel={
        \pgfkeys{/pgf/fpu=true}
        \pgfmathparse{exp(\tick)}%
        \pgfmathprintnumber[fixed relative, precision=3]{\pgfmathresult}
        \pgfkeys{/pgf/fpu=false}
      }
  }
}
    
\begin{document}
\title{\rev{Investigation of the Asymptotic Properties of Active Impedance in \revII{Large} Finite Array Antennas}}
\author{Harald Hultin, Lucas Åkerstedt and B. L. G. Jonsson
\thanks{This work was supported by the Swedish Foundation for
Strategic Research under Project ID20-0004, \revII{and the Swedish
Research Council’s Research Environment grant (SEE-6GIA
2024-06482) for research on sixth-generation wireless systems
(6G)}.}
\thanks{Harald Hultin is with Saab Surveillance, 171 54 Solna, Sweden.}
\thanks{Harald Hultin, Lucas Åkerstedt and B. L. G. Jonsson are with KTH Royal Institute of Technology, EECS,
10044 Stockholm, Sweden (e-mail: haraldhu@kth.se).}}

\maketitle

\begin{abstract}
This paper presents an improved full-wave solver for finite array antennas, and uses this solver to \rev{examine asymptotic properties of active impedance} for large \revI{regular} arrays. The improved solver is based on a preconditioning scheme that has been adapted for use on more general finite geometries, and an improved data structure. \revI{These two improvements result in a fast and stable solver with a lower memory footprint.} By investigating the active impedance \revV{of the center element in} finite arrays it is found that, even for arrays with 1000 elements, \revI{asymptotic} behavior \revI{may} differ \revV{appreciably} from infinite arrays. Tied to these properties, different predictors on how the \revI{active impedance of the} center element in the array behaves are presented. \revI{The two best predictors work very well for the wideband arrays investigated, and} may be used to decide when array approximations\revI{, such as unit cell methods,} are appropriate, rather than general statements on array size.
\end{abstract}


\section{Introduction}
\label{sec:introduction}
\bstctlcite{IEEEexample:BSTcontrol}

\IEEEPARstart{W}{hat} antenna array size is required for the center element to behave as a unit cell? This popular \cite{wu1966properties, galindo1968asymptotic, amitay1972theory, RoscoeA.J.1994Lfaa, HolterH.2002Otsr, CraeyeC.2011Aroa, BhattacharyyaArunK.2014AoFM} and challenging question is here reconsidered, with a focus on the Active Reflection Coefficient (ARC). Together with the improved Method of Moments (MoM) solver, \revIV{the results reported here provide a more nuanced understanding of when array approximations are appropriate. While a rule of thumb such as "\revI{$10\multX 10$} is sufficient" may be adequate \revV{to answer the unit cell question} in many cases, the suitability \revV{of such a rule} depends on the accuracy requirements of the intended application. The proposed predictors make it possible to relate these \revV{application} requirements to an estimated approximation error, thereby enabling a more informed decision regarding whether a given approximation is acceptable.} The results also show that asymptotic properties found for infinite arrays \revI{only weakly apply} to finite arrays\rev{, as edge effects impact the results}.

The approximation question remains important as the design of arrays is challenging. Different array applications have widely different requirements on array properties \cite{haupt2015antenna, 10930518}. Array antennas are an attractive antenna choice due a range of attractive features, including their versatility, high gain, and electrical steering \cite{haupt2015antenna}. In communication for example, it is popular to use Multiple Input Multiple Output (MIMO) antennas \cite{ikram2022road}. \rev{Normal MIMO systems usually have} arrays with tens of elements \rev{\cite{ikram2022road, lu2014overview}}. In contrast, \rev{massive MIMO systems can have more than 100 elements \cite{lu2014overview} and} radar systems use arrays that can have thousands of elements \cite{haupt2015antenna}. But even though these systems have a vastly different scale, they are still the same type of structure and originate from a similar design process.

A typical design process for an array antenna includes the following steps: design or decide upon an element which can fulfill the \revI{given requirements}, and build an array with this element. Building the \revI{array antenna} often entails simulations of the array to iterate on the element design in order to handle issues with mutual coupling, surface waves and scan loss. Both of these steps are almost exclusively performed using simulation software. In the second step, array antennas with different sizes have to be handled differently. For example, a relatively small $4\multX 4$ MIMO antenna can be simulated with a full-wave solution on a moderately equipped computer. On the other hand, arrays with thousands of elements are often unfeasible to simulate in a full-wave solver. 

\revI{A common approach to address this, is the argument that the central region of the array may behave approximately} as if the central \revI{region was} in an infinite array \cite{amitay1972theory, galindo1968asymptotic, wu1966properties}. This allows for unit cell simulations \cite{jin2008finite} to approximate the behavior of the array. Unit cells alone describe the behavior of a cell in an infinite array, but their results can be used to approximate finite arrays \cite{BhattacharyyaArunK.2014AoFM, CraeyeC.2011Aroa}. However, while efficient, they remain an approximation. Other methods such as macro basis functions \cite{maaskant2008fast, Maaskant2010Analysis, guening2025broadband, dommisse2026stabilized} and adaptive cross approximation \cite{maaskant2008fast, Maaskant2010Analysis, kurz2002adaptive, zhao2005adaptive} has been developed to simulate or approximate large arrays. \revII{An alternative to these physics or math based approximations are machine-learning assisted methods \cite{li2025macine}.} \revI{Another} class of methods that can solve the full-wave problem of the array are Toeplitz solvers \cite{brandt2024extended, cavillot2025accurate, akerstedt2024partitioning, hultin2025solver}.

A part of the results in this paper reports on the improvements of the full-wave simulations of large arrays using Toeplitz methods. We show that the preconditioning described in \secref{sec:precond} and the data structure in \secref{sec:dataStruct} \revI{reduces simulation times by a factor} 8 \revI{compared to \revII{our previous \revV{iterative} Toeplitz solver}}\cite{hultin2025solver}. \revI{The new preconditioning scheme is also well suited to the matrix structures used, due to specific properties of the Hermitian conjugate of the matrix.} Furthermore, the preconditioning strategy puts the problem on a form that allows it to be solved with the Preconditioned Conjugate Gradient method (PCG) \cite{alma99773379602456} rather than GMRES \cite{saad1986gmres}, greatly reducing the memory footprint \rev{and improving accuracy}. \revI{The improvements of the solver allows the simulation of large arrays, and to investigate }at what array size the \revII{active impedance in the} center region of the array \revII{reaches a convergent value}. \revI{The solver is compatible with very general geometries as it represents the geometry using a Rao-Wilton-Glisson (RWG) \cite{rao1982electromagnetic} mesh, and \revII{all arrays have} a finite ground plane.}

The topic of mutual coupling analysis in array antennas is a large field of research\revI{, see \cite{CraeyeC.2011Aroa} and the references therein}. The question of how large an array should be in order to be similar to an infinite array is not an easy question to answer, as the answer depends on factors such as the geometry of the array element\revI{, required accuracy,} and the scan direction \cite{amitay1972theory, JonssonB.L.G.2013AAL}. The answer \revI{may} also depend on the quantity considered \cite{wachowiak2026sizing}. Here, as in \cite{amitay1972theory, HolterH.2002Otsr}, we consider the ARC and the inter-element coupling. \revII{Inter-element coupling is often an important parameter to consider in array antenna systems \cite{10880463}.} 

There does exist some engineering thumb rules suggesting\revI{, e.g.,} that a $10\multX10$ array \revI{is} sufficient \revII{to assume similar behavior to an infinite array}. However, these rules do not provide any insight into \revIII{the accuracy of} the approximation. For example, in \cite{amitay1972theory} the infinite results are said to be reproduced "very well" when exciting a subset of $11\multX 11$ in an infinite waveguide array for non-grazing beams. In \cite{HolterH.2002Otsr} the rule-of-thumb of $10\multX10$ is generalized to $5\lambda\multX5\lambda$ and counter-examples to the $10\multX10$ rule are shown for wideband elements. The similarities to infinite arrays are given as "reasonable convergence". We propose and demonstrate how passive inter-element impedance \revI{can be used to provide} a measurement of how well ARC converges in a finite array setting. \revI{We also show that the expected behavior for passive element coupling holds well even for finite arrays, and can also be used to predict how large the error in active impedance is. These predictors \revII{also} work for arrays where the element spacing differs from $\lambda/2$, e.g., in a wideband context.}

Several other works have investigated how e.g. \revI{Floquet} theory to predicts coupling in finite arrays using results for infinite arrays. Bhattacharyya \cite{BhattacharyyaArunK.2014AoFM} shows results for waveguide openings ($11\multX11$) and patch antennas over an infinite ground plane ($3\multX3$) with good agreement, approx \SI{1.5}{\decibel} deviation in ARC. Similar methods are used in \cite{BhattacharyyaArunK.2015AAMf} to predict the excitation and active element pattern in arrays, and in \cite{lesur2018large} to model large array antennas. In addition to predicting finite array performance from infinite array, \cite{RoscoeA.J.1994Lfaa, KelleyD.F.2002Rbae} shows that the full information about coupling between elements lies in a rectangle in the $uv$-plane. In \secref{sec:predict}, it is shown that this can be reduced to a line in the $uv$-plane to predict the coupling along a row or column in the array. These types of ideas can further be extended to measurements of array antennas \cite{ctx33917433670002456}. Apart from \revI{Floquet} theory, coupling can also be estimated using e.g. reaction \cite{konforta2019approximation}. 

The paper is organized as follows. \secref{sec:theory} describes the theory used, including a short summary of the \revI{MultiLevel Fast Fourier Transform (MLFFT)} solver, the preconditioning scheme used, and how array coupling is predicted. \secref{sec:method} describes more practical aspects: calculating the blocks of the preconditioner, data arrangement and the antenna elements investigated. \revI{\secref{sec:results} presents the results, which are concluded in \secref{sec:conclusion}}.

\section{Theory}
\label{sec:theory}
To investigate the asymptotic behavior of large array requires a fast and efficient solver. The basis of the iterative solver used here are described in \secref{sec:MBT}. One of the key ingredients in a \revIII{well performing} iterative solver is the preconditioner, here described in \secref{sec:precond}. This preconditioner also enables the use of the PCG method described in \secref{sec:PCG}. The structure described in \secref{sec:MBT} also allows for simplifications in the preconditioner as shown in \secref{sec:hermprop}. Theory concerning the asymptotic behavior, antenna properties and the relations between them, is given in \secref{sec:predict}.
\subsection{The Multilevel block-Toeplitz structure}
\label{sec:MBT}
\begin{figure}[t]
    \centering
    \input{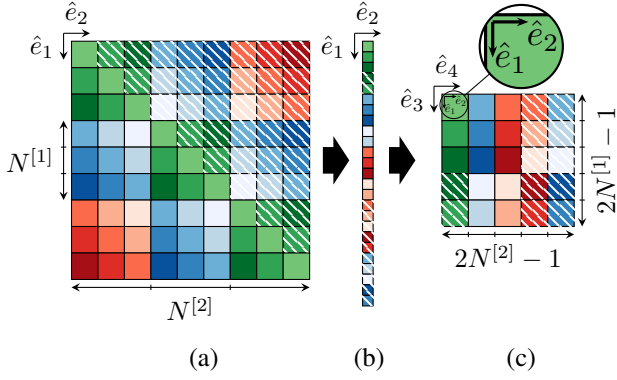}
    \caption{\revI{Visualization of data structures for a $3\multX3$ array, i.e., $\mli{N}{2}=\mli{N}{1}=3$.} \rev{(a) MBT structure, (b) MBT block vector representation\cite{hultin2025solver}, (c) MBT four dimensional representation.} \rev{Dimension $n$ is indicated by unit vectors $\hat{e}_n$. In (a) and (b), the structures are block matrices. The data inside of the blocks and the structure span the same two dimensions. In (c), the data is stored in a four dimensional matrix, where the data inside of the colored blocks still lie in dimensions 1 and 2, and the blocks lie in dimension 3 and 4.} \rev{The structure in (c) is \revV{further} described in \secref{sec:dataStruct}.} Hatched squares are the transpose (in dimensions 1 and 2) of corresponding, non-hatched squares.}
    \label{fig:ToepStruct}
\end{figure}

In \cite{akerstedt2024partitioning} it is shown that a decomposition of regular arrays results in a MoM-matrix on the form
\begin{equation}
\label{eq:Zblock}
    Z=\begin{bmatrix}
        Z_A & Z_B^T \\
        Z_B & Z_C
    \end{bmatrix}
\end{equation}
where $Z_A$ describes the interactions between antenna elements, $Z_B$ and $Z_B^T$ describes the interaction between the antenna elements and the array \revIII{margin}, and $Z_C$ describes the interactions within the \revIII{margin}. \revI{Regular arrays are here planar, fully-populated arrays consisting of identical elements that lie on a rectangular, uniform grid.} Usually the \revIII{margin} is relatively small, and $Z_A$ \revIII{undoubtedly} constitutes the majority of $Z$ \revIII{for large arrays}. \revI{With an appropriate decomposition and element indexing of a regular array, the repeated structure gives rise to a structured matrix $Z_A$.} For a 2D array, \revI{$Z_A$} will be a Multilevel Block-Toeplitz (MBT) matrix \cite{akerstedt2024partitioning}\revI{, whose structure is illustrated schematically} in \figref{fig:ToepStruct}\revI{a}.

The MBT structure can be utilized in different ways to make more efficient solver. As identical submatrices are repeated in a given pattern, it is redundant to save the whole matrix. Instead, with dedicated \rev{solver} methods, only each unique block will be \rev{stored}. While the full matrix would have a number of matrix elements that scale as $\left(\mli{N}{2}\mli{N}{1}\right)^2$, the number of unique elements scale as $\mli{N}{2}\mli{N}{1}$ for an array with $\mli{N}{1}$ rows and $\mli{N}{2}$ columns. \revI{The size of the lowest level blocks is denoted $\mli{N}{0}$. These lowest level blocks correspond to the array element mesh.} In other words, instead of having the required number of matrix elements scale quadratically with the number of antenna elements, it scales linearly.

The MLFFT with two levels $\mathcal{F}^{[2]}$ is given by \cite{hultin2025solver}:
\begin{equation}
\label{eq:F2}
    \mathcal{F}^{[2]} = \mathcal{F}_{N^{[2]}} \otimes \left(\mathcal{F}_{N^{[1]}} \otimes \rev{\id}_{N^{[0]}}\right),
\end{equation}
\revI{where $\id_{\mli{N}{0}}$ is the identity matrix with dimension $\mli{N}{0}$. Using the MLFFT,} the multiplication $ V_A =Z_AI_A$ can be accelerated by \revI{embedding the MBT matrix in a multilevel block circulant matrix $C^{[2]}(Z_A)$:}
\begin{multline}   
\label{eq:multL}
    C^{[2]}(Z_A)I_A^{[L]} = \\ \left(\mathcal{F}^{[2]}\right)^{-1} \text{diag}_{N^{[0]}}\left(\mathcal{F}^{[2]}C^{[2]}_1(\revII{Z_A})\right)\mathcal{F}^{[2]}I_A^{[2]}=V_A^{[2]} \\
    I^{[L]},V^{[L]} \in \mathbb{C}^{(2\mli{N}{2}-1)(2\mli{N}{1}-1)\mli{N}{0}},
\end{multline}
where $\text{diag}_{N^{[0]}}$ denotes a block diagonal matrix with block size $N^{[0]}$ and vectors $I_A^{[2]}$ and $V^{[2]}$ are padded versions of $I_A$ and $V_A$. \revI{While $C^{[2]}(Z_A)$ is larger than $Z_A$, the multiplication via MLFFT only requires the first column in the circulant block matrix, denoted by $C_1(\revII{Z_A})$ and illustrated in \figref{fig:ToepStruct}b.} Thus, by utilizing the MBT structure illustrated in \figref{fig:ToepStruct}a, and arranging the unique elements as the first column in \revI{the circulant block matrix} in \figref{fig:ToepStruct}b, the multiplication with the MBT matrix is greatly accelerated \cite{hultin2025solver}.

\subsection{Block diagonal preconditioner}
\label{sec:precond}
\revII{Preconditioners are often used together with iterative solvers to accelerate convergence \cite{carson2024towards}. Since the spectral properties of MoM matrices can lead to slow convergence of Krylov-subspace methods \cite{carson2024towards}, we employ a preconditioning strategy inspired by \cite{cavillot2025accurate}. There, the preconditioner $P_T=D$, is defined as the block-diagonal part of $Z^HZ$, where $Z$ mainly consist of a Toeplitz-block-block-Toeplitz structure. The structure comes from the use of a rectangular basis function in the x- and y-directions, resulting in relatively small innermost block size. Here $(\cdot)^H$ denote the Hermitian conjugate. The resulting preconditioned system can be written as}
\begin{equation}
    \label{eq:precondTBBT}
    \revII{P_T^{-1} Z^H ZI=P_T^{-1} Z^H V}
\end{equation}

\revII{Since $Z^HZ$ is self-adjoint, its spectrum is on the real-axis. Note that $P_T$ is defined through a partitioning based on geometric rows, and is closely related to the preconditioner $P_Z$ introduced in \cite{hultin2025solver}.}

\revII{To generalize this preconditioning strategy to fully three-dimensional geometries discretized with RWG basis functions \cite{rao1982electromagnetic}, we exploit the matrix partitioning described in \cite{akerstedt2024partitioning}. This partitioning yields an impedance matrix of the form given in \eqref{eq:Zblock}, where the submatrix $Z_A$ possesses an MBT structure. In contrast to the formulation in \cite{cavillot2019efficient}, the impedance matrix in \eqref{eq:Zblock} also contains the border-border coupling matrix $Z_C$ and the border-interior coupling matrix $Z_B$, both of which must be incorporated into the preconditioning algorithm. The resulting matrix takes the form }

\begin{equation}
\label{eq:Z2}
    \begin{split}
        \begin{bmatrix}
            Z_A^H & Z_B^H \\
            Z_B^* & Z_C^H
        \end{bmatrix}
        \begin{bmatrix}
            Z_A & Z_B^T \\
            Z_B & Z_C
        \end{bmatrix}
        \begin{bmatrix}
            I_A \\
            I_C
        \end{bmatrix}
        =
        \begin{bmatrix}
            Z_A^H & Z_B^H \\
            Z_B^* & Z_C^H
        \end{bmatrix}
        &\begin{bmatrix}
            V_A \\
            V_C
        \end{bmatrix}
        \\
        =
        \begin{bmatrix}
            Z_A^H Z_A+Z_B^H Z_B & Z_A^H Z_B^T+Z_B^H Z_C \\
            Z_B^* Z_A +Z_C^H Z_B & Z_B^* \revII{Z_B^T}+Z_C^H Z_C
        \end{bmatrix}
        &\begin{bmatrix}
            I_A \\
            I_C
        \end{bmatrix},
    \end{split}
\end{equation}
\revII{where $(\cdot)^*$ is the element-wise complex conjugate. Since $Z_C$ is a relatively small and $Z_B$ is low rank, $Z^HZ$ can evaluated effectively, especially when the non-MBT margin is thin. However, the  reformulation \eqref{eq:Z2} may increase the condition number of the system, making preconditioning even more important. Therefore the preconditioner $P$ is defined as the block-diagonal part, and its construction is described in \secref{sec:precondcalc}.}

\revII{It should be emphasized that $P$ preconditions the normal-equation system $Z^HZI = Z^HV$. In contrast the term ”preconditioning strategy” refers to the complete procedure of application of $Z^H$ followed by the preconditioner $P$ to the original system $ZI=V$. The effectiveness of this strategy is demonstrated in \secref{sec:results} for both planar and fully three-dimensional elements.}

\subsection{Preconditioned conjugate gradient method}
\label{sec:PCG}
As $Z^HZ$ is positive definite \cite{alma99286799402456}, this problem can be solved using conjugate gradient methods \cite{alma99773379602456}. In this paper, the PCG method is used. This method should be similar in speed to GMRES, but importantly it only saves a small number of vectors \rev{with the same size as $V$}. \rev{So while GMRES memory grows linearly in problem size and number of iterations, PCG only grows linearly in problem size\cite{alma99773379602456}.} This leads to \revI{both less memory usage, and more control over memory usage,} in most cases. \rev{Using PCG} improves the results in \cite{hultin2025solver} that used GMRES and partly suffered from the large dimension of the Krylov subspace associated with GMRES \cite{saad1986gmres}.

\subsection{Commutative Hermitian \revI{conjugate} property of block Toeplitz $Z$ in MLFFT methods}
\label{sec:hermprop}
An essential part of an efficient solution scheme of $ZI=V$ is fast matrix multiplication, \revI{here using} MLFFT. One weakness of the proposed preconditioning strategy in \secref{sec:precond} is that, for an arbitrary matrix $G$ the Hermitian conjugate in general is not commutative with the Fourier transform, and by extension not with the MLFFT: $\mli{\mathcal{F}}{2}(G^H)\neq \left(\mli{\mathcal{F}}{2}(G)\right)^H$. Thus the precondition strategy would require storage of $\mli{\mathcal{F}}{2}(\revII{Z_A^H})$. \rev{While $\mli{\mathcal{F}}{2}(\revII{Z_A^H})$ can be stored efficiently like $\mli{\mathcal{F}}{2}(\revII{Z_A})$, it still \revI{requires the same amount of memory as $\mli{\mathcal{F}}{2}(\revII{Z_A})$}. This is not detrimental to the solver, but inefficient.} \revIV{However, as shown in Appendix~\ref{app:herm}, the MLFFT and the Hermitian conjugate commute for MBT impedance matrices:}
\begin{equation}
    \label{eq:multCom}
    \mathcal{F}^{[2]}(C_1(Z_A^H))=[\mathcal{F}^{[2]}(C_1(Z_A))]^H.
\end{equation}
Thus, only $\mathcal{F}^{[2]}(C_1(Z_A))$ has to be calculated, and the Hermitian \revI{conjugate} can be applied during multiplication.

Therefore, $Z^HZv$ can be calculated by only saving $\mathcal{F}^{[1]}(C_1((\mli{T}{1})))$ in memory and applying the conjugate at the time of multiplication. 

\revI{Recall that $Z_A$ is a 2-level matrix for a 2\revI{D} array, but this proof is performed on 1-level for brevity. As a 2-level MLFFT can be seen as applying a 1-level MLFFT twice,} the 1-level result \revI{above can be} generalized to the 2-level structure by replacing the blocks $t_n$ in \eqref{eq:scalarC} with $\mathcal{F}^{[1]}(C(\mli{T}{1}_m))$, where $\mli{T}{1}_m$ are 1-level block-Toeplitz matrices.

To solve the excitation of the array antenna, one considers a linear equation problem on the form
\begin{equation}
\label{eq:eqProb}
    ZI=V,
\end{equation}
where $I$ are the currents and $V$ is the excitation voltage \rev{on the mesh}. The variables are vectors, \rev{$I,\:V\in \mathbb{C}^{\mli{N}{2}\mli{N}{1}\mli{N}{0}}$}, \revI{when} solving for a single excitation \revI{voltage $V$}. \rev{But they can also be matrices, e.g. when solving for all Embedded Element Patterns (EEPs) simultaneously, as in this publication and \cite{hultin2025solver}. For an array with $\mli{N}{2}\mli{N}{1}$ elements and where each element has $F$ feeds to find an EEP for, $I,\:V \in \mathbb{C}^{\mli{N}{2}\mli{N}{1}\mli{N}{0}\multX \mli{N}{2}\mli{N}{1}F}$. Here, $F=1$.}

\subsection{Predicting inter-element coupling}
\label{sec:predict}
Coupling between array elements is a physical quantity that describes how some excitation of one antenna effects another antenna. This coupling can be described using different parameters\revI{. In} this publication the focus is on active parameters between antenna elements in an array\revI{, i.e.,} those that the array exhibit given some excitation \revI{$V$}. The active impedance \revI{$Z_{a,i}$} seen at antenna port $i$, given \rev{that the array} \revI{has} \rev{a set of port currents \revI{$I_{p,j}, \: j=1,\dots, N$} at each port $j$}, is given by
\begin{equation}
\label{eq:actZ}
    \revI{Z_{a,i} = \frac{1}{I_{p,i}} \sum_{j=1}^{N} Z_{p,ij} \, I_{p,j},}
\end{equation}
where \revI{$Z_{p,ij}$} is the passive impedance between ports $i$ and $j$. This equation can also be put in matrix form as
\begin{equation}
\label{eq:actZmat}
    \revI{Z_a = \text{diag}(I_p)^{-1} Z_p I_p},
\end{equation}
where \revI{$Z_a$ and $I_p$ are vectors with elements $Z_{a,i}$ and $I_{p,j}$} respectively, and $Z_p$ is a matrix with elements $Z_{p,ij}$.  The passive impedance relates current and voltages \rev{$\portV{i}$ between ports as}
\begin{equation}
\label{eq:passZ}
    \rev{Z_{\revIII{p},ij} = \frac{\portV{i}}{\portI{j}}, \: \portI{k}=0 \: \text{for} \: k \neq j.}
\end{equation}
\revI{Note that $Z_{p,ii}$ will be referred to as the self impedance of element $i$, and $Z_{p,ij}$ as the mutual impedance between elements $i$ and $j$.}

A closely related quantity to active impedance is the Active Reflection Coefficient (ARC) \revI{$\Gamma_{a,i}$}:
\begin{equation}
\label{eq:ARC}
    \revI{\Gamma_{a,i} = \frac{Z_{a,i} - Z_0}{Z_{a,i} + Z_0}},
\end{equation}
\revI{where $Z_0$ is the characteristic impedance. In this publication, $Z_0$ is real valued. ARC} describes the outgoing voltage wave of port $i$ for the given excitation, an active analog to the reflection coefficient $\Gamma_i$.

To be able to calculate either the active impedance or ARC of an array for any excitation, each element excitation must be simulated separately to find each passive impedance in \eqref{eq:passZ}. This quickly becomes computationally expensive as the number of array elements grow. \rev{Approximations such as} unit cell solutions are a \revI{cheaper} alternative to calculate for example the active impedance of a unit cell. \revI{Denoting an approximation} \revI{of $Z_{a,c}$ as $\tilde{Z}_{a}$, where $c$ indicates the center element, introduce the relative error} \revI{$\xi$}:
\begin{equation}
\label{eq:xi}
    \tilde{Z}_{a} = Z_{a,c}(1+\revI{\xi}).
\end{equation}

\rev{If an approximation \revI{$\tilde{Z}_{a}$} is used instead of the true solution,} \revI{it is important to know the size of the relative error $\xi$. Here, we introduce two different approximations to $\xi$: $\xi_Z$ and $\xi_S$,} \rev{that are derived either from the \revV{impedance \eqref{eq:passZ}}, or the expected decay of the scattering matrix $S$ respectively.}

\rev{To find $\xi_Z$, consider} an array with a uniform amplitude excitation \revI{$V_p$, i.e. each antenna element has the same excitation} \revI{voltage}, where the center element with index $c$ has the active impedance $Z_{A,c}$ and all elements have a similar self impedance \revI{$Z_{p,ii}$, $|\portI{j}|\approx|\portI{i}|\: \forall i,j$}. If one element \revI{$b$} is added to the array, it can be observed that the new active impedance \revI{$Z'_{a,c}$} is given by
\begin{equation}
    \revI{Z'_{a,c} \approx Z_{a,c} + Z_{p,cb}e^{\text{j}\revI{(\angle \portI{b}-\angle \portI{c})}}}.
\end{equation}
Using the triangle inequality, this gives that
\begin{equation}
    \revI{|Z'_{a,c}| \approx |Z_{a,c} + Z_{a,cb}e^{\text{j}\revI{(\angle \portI{b}-\angle \portI{c})}}| \leq |Z_{a,c}| + |Z_{a,cb}|.}
\end{equation}
When adding an element to the array, the change in active impedance of the center element is expected to be smaller than the passive impedance between the center element and the new element. Extending this to the case where the array grows in each direction, i.e. by adding a set \revI{$E_B$ consisting of} $N_B$ new border elements \rev{such that the size increases from $N\multX N$ to $(N+2)\multX (N+2)$}, a similar argument could be made that the approximate change in active impedance \revI{$\Delta Z_{a,c}$} would be limited as
\begin{equation}
\label{eq:badpred}
    \revI{|\Delta Z_{a,c}| \leq \sum_{n\in \revI{E_B}} |Z_{p,cn}| = N_B|Z_{p,cB}|,}
\end{equation}
\revI{where $|Z_{p,cB}|$ is the average amplitude of $Z_{p,cn}, \,n\in E_B$} However while this expression gives an \rev{upper bound}, it is in practice \rev{very} conservative. \rev{For example, in \secref{sec:results}, relative errors tend to lie between \SIrange[]{E-2}{E-3}{} for arrays with more than a hundred elements ($N>10$). This means that the estimated error using \eqref{eq:badpred} would be on the order of $100\%$.} 

\revIV{What can be deduced from \eqref{eq:badpred} is that the error introduced when extending the array is related to the set of mutual impedances between the center element and the newly added border elements. However, as the impedance terms are complex-valued, the bound is highly conservative since it assumes that all contributions add coherently. In practice, the contributions exhibit varying phases, resulting in significant cancellation. The magnitude of the error is therefore typically governed not by the sum of all border couplings, but by the largest individual coupling introduced by the extension. An estimator of the error can then be defined by considering the maximum impedance introduced:}
\begin{equation}
    \xi_Z=\frac{\max\limits_{j \in \revI{E_B} } \left| \Zp{cj} \right|}{\left| \Zp{cc} \right|}.
\end{equation}
\revIV{Furthermore, the mutual impedance generally decreases with increasing element separation. Consequently, the border elements closest to the center element are expected to dominate the perturbation, while the remaining border elements contribute increasingly smaller terms. Motivated by this observation, the predictor $\xi_Z$ is defined as the largest normalized mutual impedance introduced by the border, which can be interpreted as a first-order estimate of the scale of the perturbation for the active impedance of the center element:} 
\begin{equation}
\label{eq:predZ}
    \xi_Z\approx\frac{\left| \Zp{cj} \right|}{\left| \Zp{cc} \right|}, \: \text{\revII{$j\in E_B,$ such that $j$ closest to $c$}}.
\end{equation}

\revIV{The predictor \eqref{eq:predZ} should not be interpreted as an exact bound derived from \eqref{eq:badpred}. Rather, it is a heuristic predictor motivated by two observations: the error must be related to the couplings introduced by the newly added border elements, and that these couplings decrease with distance such that the nearest border elements are expected to dominate. The effectiveness of the predictor is therefore empirical and evaluated through the numerical results of \secref{sec:results}.}

\revII{Solving for a single passive impedance is relatively inexpensive computationally. However, an even cheaper predictor can be obtained by considering the expected asymptotic behavior of the array described in \cite{amitay1972theory}.}

\revI{The relevant results \revII{in \cite{amitay1972theory}} build on two earlier publications \cite{wu1966properties, galindo1968asymptotic}. In \cite{wu1966properties}, the asymptotic properties of an infinite array consisting of thin walled waveguide openings is studied, and in \cite{galindo1968asymptotic} a similar study is performed for infinite planar arrays. For the waveguide openings, the problem can be reduced down to a 1D problem of infinite parallel plate waveguides. For these waveguides the coupling coefficients along the plane perpendicular to the plates are found to decay $\propto \elDist ^{-3/2}$ \cite{amitay1972theory, wu1966properties}, where $\elDist$ is element distance, i.e. the number of elements between the two considered elements or columns. For the planar array, the problem is evaluated as a 2D problem, and the decay is found to be $\propto \elDist^{-2}$ \cite{amitay1972theory,galindo1968asymptotic}. The 2D analysis is similar to other studies using Floquet theory \cite{BhattacharyyaArunK.2014AoFM}. An attractive aspect of the study in \cite{wu1966properties}, is the fact that it reduces down to a 1D problem. And given that the very general 2D analysis in \cite{galindo1968asymptotic} seems to fit well with finite arrays as in \cite{BhattacharyyaArunK.2014AoFM}, we will here apply the 1D theory of \cite{wu1966properties} to non-ideal, finite arrays to see how well it works as a predictor.} 

For \revII{a planar, rectangular} array antenna \rev{with a uniform amplitude excitation in voltage,} $\Gamma_{\revI{a},[p,q]}(\psi_x, \psi_y)$ of the element on row $p$ and column $q$ is given by 
\begin{equation}
    \Gamma_{\revI{a},[p,q]}(\psi_x, \psi_y) = \sum_{m,n}S_{[p,q][m,n]}\e^{\ju (m\psi_y+n\psi_x)},
\end{equation}
where the sum is taken over all rows $m$ and columns $n$, $S_{[p,q][m,n]}$ is the scattering parameter from the element on row $m$, column $n$, to the element on row $p$, column $q$, and $\psi_x,\ \psi_y$ are the incremental phases between elements in the $x, \ y$ direction (along rows and columns respectively). For an infinite array, all elements have identical $\Gamma_{\revI{a}}$ and the indices $p,\ q$ can be omitted. \rev{Furthermore, considering only the behavior along a row $p=0$, while also only \revI{scanning along the E-plane ($\psi_y=0$)}, the expression can be simplified further to}
\begin{equation}
    \Gamma_{\revI{a}}(\psi_x) = \revI{\sum_{m,n}S_{[m,n]}\e^{\ju n\psi_x}=\sum_{n}\mathcal{C}_{n}\e^{\ju n\psi_x}},
\end{equation}
\revI{where $\mathcal{C}_n$ is the scattering between two columns in the array spaced $n$ elements apart, analog to the scattering between parallel plates in \cite{wu1966properties, amitay1972theory}.}

In this way, \rev{\cite{amitay1972theory, wu1966properties} shows} that, for an infinite array of ideal waveguide openings the coupling $\colS{n}$ between \revI{columns} is given by Fourier coefficients of the active reflection coefficient:

\begin{equation}
\label{eq:amitayPred}
    \colS{n} = \frac{1}{2\pi}\int_{-\pi}^\pi \Gamma_{\revI{a}}(\psi)\e^{-\ju n\psi}\mathrm{d}\psi,
\end{equation}
where $\psi$ is the incremental phase between elements along the row or column. \revI{Note that along a column or row, $\elDist=n$.} Similar results are shown in \cite{BhattacharyyaArunK.2014AoFM}, where the unit cell behavior \rev{in a fully excited array} is used to predict the coupling to nearby elements. \rev{This allows for calculating coupling to all elements, but is more expensive compared to \eqref{eq:amitayPred}.} 

One of the key results from \cite{galindo1968asymptotic, amitay1972theory, wu1966properties} is the asymptotic behavior of the coupling coefficients. It is shown that \revI{$|S_{cn}|\propto \elDist^{-2}$. Further,} it is stated that "The asymptotic dependence on distance of the coupling coefficients seems to be universal for all phased arrays" \cite{amitay1972theory}. \revI{The} theory is proven for infinite arrays with either a finite or infinite amount of elements excited. It is not shown for finite arrays. Proving these properties for a finite array is likely not possible. However, applying \eqref{eq:amitayPred} to finite arrays with different properties gives an insight in to how well the theory extends to the general finite case. As \rev{Wu \revII{and Galindo} \cite{wu1966properties}} states, the asymptotic slope can be tied to the discontinuity in the directional derivative of the active reflection coefficient when either a grazing lobe appears, or the main lobe disappears. This discontinuity is not present in the finite arrays examples \revI{described in \secref{sec:antennas}}\rev{, as the sharpness in part comes from that all energy goes back through the ports. In most finite antennas, except for some extreme outliers, some energy will radiate out. For example, if a critical steering angle is reached such that the power propagates along the surface of the array, it will scatter at the edge of the array}.

The active reflection coefficient is often evaluated over a scan range, \revI{a range of directions} such as a cut in the $uv$-plane. But it is important to note that the phase in \eqref{eq:amitayPred} is evaluated in a cut plane, but not necessarily over the full range. Given that $\theta_0$ is the angle from boresight the beam should be directed at, the excitation \rev{voltage} of each element is given by
\begin{equation}
    \rev{V}_{m,n} = \revI{\e^{\ju n\kappa d_xsin\theta_0} = \e^{\ju n\psi}},
\end{equation}
where \revI{$\kappa$} is the wavenumber and $d_x$ is the element size in the scan direction.

For ARC evaluated over a scan range in $\theta$, \revI{$\hat{\Gamma}_{\revII{a}}(\theta)$}, the phase is a function of the scan angle $\varphi(\theta)$. With the variable substitution $\hat{\Gamma}_{\revII{a}}(\theta)=\Gamma_{\revII{a}}(\revI{\varphi}(\theta))$, \eqref{eq:amitayPred} becomes
\begin{equation}
\label{eq:amitayPredTheta}
    \colS{n} = \frac{1}{2\pi}\int_{-\pi}^\pi \hat{\Gamma}_{\revII{a}}\left(\sin^{-1}\left(\frac{\psi}{kd}\right)\right)\e^{-\ju n\psi}\mathrm{d}\psi.
\end{equation}
As is evident, the scan range to integrate over depends on the element size. For an \revI{element spacing $d=\lambda/2$}, this results in integrating over a scan range between $\theta=\pm\ang{90}$, while e.g. a $0.6\lambda$ spacing results in integrating over a scan range between $\theta=\pm\ang{56}$. \revI{This integration region is defined from the relationship between scan angle and progressive phase: $\psi=kd\sin{\theta}$, and is not to be confused with, e.g., the grating-lobe free region which would be $\theta\approx\pm\ang{42}$ for $0.6\lambda$ spacing.} Thus, evaluating the ARC of the center element over a limited cut can give a prediction of how the scattering parameters behave asymptotically.

The knowledge of how $S$ behaves asymptotically can further be used to predict the behavior of $\revI{Z_p}$ and $\revI{Z_a}$. For an array where every port has the reference impedance $Z_0$, the conversion between $S$ and $\revI{Z_p}$ is given by \cite{alma99562318102456} as:
\begin{equation}
    \label{eq:S2Z}
    \revI{Z_p}=Z_0(\id+S)(\id-S)^{-1},
\end{equation}
where $\id$ is the identity matrix. \revIV{As shown in Appendix~\ref{app:scaling}, it is expected that the addition of weak entries in $S$ with a ceratin scaling, should produce corresponding entries in $Z_p$ with the same scaling. As the error in $Z_a$ is expected to be proportional to the weak entries in $Z_p$, this provides another prediction: that $\xi$ should decay $\propto \elDist^{-2}$.}

\rev{In theory, $\colS{n}$ in \eqref{eq:amitayPredTheta}} \revI{of a planar array} \rev{should decay as $\elDist^{-2}$ for large $\elDist$}\revI{\cite{amitay1972theory, galindo1968asymptotic}. Thus, a }\rev{first order approximation of $\Zp{cj}$, and thereby \revI{$\xi$}, is that it also scales as $\elDist^{-2}$. In practice, to find $\xi_S$ an idealized $S$-matrix $S_{\text{ideal}}$ can be built from the expected convergent behavior and inserted into \eqref{eq:S2Z} to give an approximation of how $\Zp{cj}$ and \revI{$\xi$} behave for growing arrays. This matrix $S_{\text{ideal}}$ is defined to have the same $S_{nn}$ as the \revI{center element in a large array}. Off-center coupling coefficient decay \revII{monotonically} as $\elDist^{-2}$, as if in an infinite array, and also include a free-space phase-shift. The deviation $\xi_S$ via converting $S_{\text{ideal}}$ using \eqref{eq:S2Z} and finding the expected deviation as in \eqref{eq:predZ}. This results in the expression}
\begin{multline}
\label{eq:predS}
    \rev{\xi_S=\frac{\left| \left[Z_0(\id+S_{\text{ideal}})(\id-S_{\text{ideal}})^{-1}\right]_{cj} \right|}{\left| \Zp{cc} \right|},}\\ \text{\revII{$j\in E_B,$ such that $j$ closest to $c$}}.
\end{multline}

\section{Method}
\label{sec:method}
\subsection{Calculating the preconditioner for $Z^HZ$}
\label{sec:precondcalc}
To solve \eqref{eq:Z2} efficiently using an iterative method requires a preconditioner for $Z^HZ$. \revII{Here, the preconditioner will be divided into two parts, and these will correspond to the two blocks along the diagonal of the RHS in \eqref{eq:Z2}.} Using the same principles as the $P_K$ in \cite{hultin2025solver}, \revII{these two parts will consist of} the block diagonal of $\revII{Z_A^HZ_A}$ on the lowest level, i.e. the smallest squares in \figref{fig:ToepStruct} \rev{with size $\mli{N}{0}\multX\mli{N}{0}$} \revI{together with the block corresponding to $Z_C$ in \eqref{eq:Zblock}}. Unlike the preconditioner from \cite{hultin2025solver} \revII{defined on the identical blocks on the block diagonal of $Z_A$}, \revI{the blocks along the \revII{block diagonal of $Z^H_AZ_A$} are largely unique}.

\revII{As can be seen in \eqref{eq:Z2}, the upper left part is $Z_A^HZ_A+Z_B^HZ_B$. But} as $Z_B$ is low rank, \revII{and for efficiency,} the first part of the preconditioner is defined \revII{only} from the block diagonal of $Z_A^HZ_A$. Looking at the structure in \figref{fig:ToepStruct}, it is apparent that each block along the diagonal will be on the form
\begin{equation}
\label{eq:precondA}
    \revI{[P_A]_{i,i}}=\sum _{j=1}^{\mli{N}{2}\mli{N}{1}}\revI{[Z^H_A]_{j,i}[Z_A]_{j,i}}=\sum _{j=1}^{\mli{N}{2}\mli{N}{1}}\revI{[Z_A^*]_{i,j}[Z_A]_{j,i}}.
\end{equation}
Here, the preconditioner is found by precomputing all \revII{$[Z_A^H]_{j,i}[Z_A]_{j,i}$} combinations, and then summing the subsets of these combinations for each $i$. \revIV{This is also illustrated in \figref{fig:PAcalc}, where it can be seen that each block in $P_A$ is given by a sum of matrix multiplications. Each multiplication is with the conjugate or Hermitian of a block in $Z_A$, with itself or the transpose of itself.} An alternate approach is to consider how $P_{i,i}$ changes when increasing $i$ by one \cite{cavillot2025accurate}. Given the relatively small size of the blocks in \eqref{eq:precond}, the precomputation time is negligible compare to the solution time. \revIV{While the temporary saving of the $[Z_A^H]_{j,i}[Z_A]_{j,i}$ combination requires memory, these $\mli{N}{2}\mli{N}{1}$ submatrices will be cleared before calculating the MLFFT of $Z_A$. As the MLFFT of $Z_A$ also consist of $\mli{N}{2}\mli{N}{1}$ submatrices, the temporary allocation is not a bottleneck for memory.}

\begin{figure}
    \centering
    \input{precondconstruct}
    \caption{\revIV{Illustration of calculating $P_A$ (blocks 1-9) in \eqref{eq:precondA} using the example in \figref{fig:ToepStruct}. The diagonal hatches that slope from top left are transposes, diagonal hatches that slope from top right are element-wise complex conjugate, and cross-hatch is the Hermitian conjugate.}}
    \label{fig:PAcalc}
\end{figure}

\revI{The second part of the preconditioner is the inverse of the corresponding, small matrix $P_C=Z_B^* \revII{Z_B^T}+Z_C^H Z_C$. The complete preconditioner $P$ is thus given by}
\begin{equation}
    \label{eq:precond}
    \revI{    P=\begin{bmatrix}
        P_A & 0 \\
        0 & P_C
    \end{bmatrix}.}
\end{equation}

\revI{As $P$ is a Hermitian matrix it can be Cholesky factorized \cite{alma99286799402456} as $P=LL^H$, where $L$ is a lower triangular matrix. Similar to an LU factorization, this factorization allows for a fast application of $P^{-1}$, with the added benefit of not having to save the upper triangular matrix $U$ explicitly. For GMRES, this is used to apply $P$ as $P^{-1}Z^HZ\revII{I}=P^{-1}Z^H\revII{V}$. For PCG, $P$ is applied as $P^{-\frac{1}{2}}Z^HZP^{-\frac{H}{2}}W=P^{-\frac{1}{2}}Z^H\revII{V}$, where $W=P^{\frac{H}{2}}\revII{I}$ \revII{and $P^{-\frac{1}{2}}$ is the principal square root of $P^{-1}$}. Preconditioners are applied using the corresponding built in functions in MATLAB. Note that $P^{-1}Z^HZ$ and $P^{-\frac{1}{2}}Z^HZP^{-\frac{H}{2}}$ will have identical spectra $\sigma$, as $\sigma(AB)=\sigma(BA)$ and $P$ is self-adjoint \cite{alma99286799402456}.}

\subsection{Data arrangement for fast MLFFT}
\label{sec:dataStruct}
The MLFFT implemented as in \cite{hultin2025solver} works \revI{well}, but suffers performance wise from having to jump between different parts of the array in \figref{fig:ToepStruct}\revI{b}, to perform the two FFTs. \rev{This can be seen in the code in the appendix of \cite{hultin2025solver}. To perform the MLFFT on \figref{fig:ToepStruct}\revI{b}, each $n$:th row of every block must be extracted to a new matrix $M$. An FFT is then performed on $M$, the resulting FFT is redistributed back to its rows, and the procedure is repeated for every block on the levels, i.e. $\mli{N}{2}\mli{N}{1}$ times. Moving data from separate rows in the data structure to $M$ takes time, and so does jumps when indexing the memory of the separate rows.} The performance of the MLFFT can be improved by instead structuring the data in as a 4-dimensional array as in \figref{fig:ToepStruct}\revI{c}. 

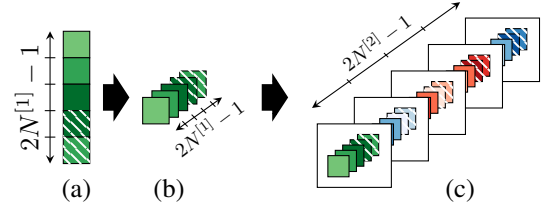
\begin{figure}
    \centering
    \begin{tikzpicture}[scale=0.35, 
Hatch/.style={postaction={
pattern={
Lines[angle=-45, distance=1mm,  line width=0.25mm]
},    pattern color=white,
}},
Hatch2/.style={postaction={
pattern={
Lines[angle=-45, distance=1mm,  line width=0.25mm]
},    pattern color=lightgray,
}},
spy using outlines={circle, magnification=3, connect spies}]

\def\cellsize{1} 
\usetikzlibrary{patterns}
\usetikzlibrary{patterns.meta}
\usetikzlibrary{arrows, arrows.meta}

\definecolor{Cell1h1}{RGB}{116,196,118}
\definecolor{Cell1h2}{RGB}{49,163,84}
\definecolor{Cell1h3}{RGB}{0,109,44}
\colorlet{Cell1h4}{Cell1h2}
\colorlet{Cell1h5}{Cell1h3}

\definecolor{Cell4h1}{RGB}{8,81,156}
\definecolor{Cell4h2}{RGB}{49,130,189}
\definecolor{Cell4h3}{RGB}{107,174,214}
\definecolor{Cell4h4}{RGB}{189,215,231}
\definecolor{Cell4h5}{RGB}{239,243,255}

\colorlet{Cell2h1}{Cell4h5}
\colorlet{Cell2h2}{Cell4h4}
\colorlet{Cell2h3}{Cell4h3}
\colorlet{Cell2h4}{Cell4h2}
\colorlet{Cell2h5}{Cell4h1}

\definecolor{Cell5h1}{RGB}{165,15,21}
\definecolor{Cell5h2}{RGB}{222,45,38}
\definecolor{Cell5h3}{RGB}{251,106,74}
\definecolor{Cell5h4}{RGB}{252,174,145}
\definecolor{Cell5h5}{RGB}{254,229,217}

\colorlet{Cell3h1}{Cell5h5}
\colorlet{Cell3h2}{Cell5h4}
\colorlet{Cell3h3}{Cell5h3}
\colorlet{Cell3h4}{Cell5h2}
\colorlet{Cell3h5}{Cell5h1}

\newcommand{\BoundingBox}[3]{%

\fill[gray!20,opacity=.15]
(0,0,0) --
(#1,0,0) --
(#1,-#2,0) --
(0,-#2,0) -- cycle;

\fill[gray!15,opacity=.10]
(0,0,0) --
(#1,0,0) --
(#1,0,#3) --
(0,0,#3) -- cycle;

\fill[gray!10,opacity=.10]
(#1,0,0) --
(#1,-#2,0) --
(#1,-#2,#3) --
(#1,0,#3) -- cycle;

\draw
(0,0,0) -- (#1,0,0) -- (#1,-#2,0) -- (0,-#2,0) -- cycle
(0,0,#3) -- (#1,0,#3) -- (#1,-#2,#3) -- (0,-#2,#3) -- cycle
(0,0,0) -- (0,0,#3)
(#1,0,0) -- (#1,0,#3)
(#1,-#2,0) -- (#1,-#2,#3)
(0,-#2,0) -- (0,-#2,#3);
}

\node[] at (0.5,-6) {(a)};
\node[] at (4,-6) {(b)};
\node[] at (15,-6) {(c)};

\begin{scope}[scale=1, xshift=0*\cellsize cm, yshift=0*\cellsize cm]

    \draw[fill=Cell1h1, draw=black] (0*\cellsize,0) rectangle ++(\cellsize,-\cellsize);
    \draw[fill=Cell1h2, draw=black] (0*\cellsize,-1*\cellsize) rectangle ++(\cellsize,-\cellsize);
    \draw[fill=Cell1h3, draw=black] (0*\cellsize,-2*\cellsize) rectangle ++(\cellsize,-\cellsize);
    \draw[fill=Cell1h5, draw=black, Hatch] (0*\cellsize,-3*\cellsize) rectangle ++(\cellsize,-\cellsize);
    \draw[fill=Cell1h4, draw=black, Hatch] (0*\cellsize,-4*\cellsize) rectangle ++(\cellsize,-\cellsize);

    \draw[draw=black, <->, >=stealth] (-0.5,0) -- ++(0,-5*\cellsize) node[midway, above, rotate=90] {\revI{$2N^{[1]}-1$}};

    \foreach \x in {1,2,3,4}
        \draw (-0.7,-\x*\cellsize) -- (-0.3,-\x*\cellsize);


\end{scope}

\draw[-{Triangle[width=18pt,length=6pt]}, line width=10pt](1.5,-2.5) -- (2.5, -2.5);

\begin{scope}[
    x={(1cm,0cm)},
    y={(0cm,-1cm)},
    z={(0.35cm,0.25cm)},
    xshift=3*\cellsize cm, yshift=-3.5*\cellsize cm
]


    \draw[fill=Cell1h4, draw=black, Hatch] (0*\cellsize,0,4*\cellsize) rectangle ++(\cellsize,-\cellsize,0);
    \draw[fill=Cell1h5, draw=black, Hatch] (0*\cellsize,0,3*\cellsize) rectangle ++(\cellsize,-\cellsize,0);
    \draw[fill=Cell1h3, draw=black] (0*\cellsize,0,2*\cellsize) rectangle ++(\cellsize,-\cellsize,0);
    \draw[fill=Cell1h2, draw=black] (0*\cellsize,0,1*\cellsize) rectangle ++(\cellsize,-\cellsize,0);
    \draw[fill=Cell1h1, draw=black] (0*\cellsize,0,0) rectangle ++(\cellsize,-\cellsize,0);

    \draw[draw=black, <->, >=stealth] (1.25,0.25,0) -- ++(0,0,5*\cellsize) node[midway, below, rotate=37, scale=0.75] {\revI{$2N^{[1]}-1$}};

    \foreach \x in {1,2,3,4}
        \draw (1.15,0.15,\x*\cellsize) -- (1.35,0.35,\x*\cellsize);

\end{scope}

\draw[-{Triangle[width=18pt,length=6pt]}, line width=10pt](7.5,-2.5) -- (8.5, -2.5);

\begin{scope}[
    x={(1cm,0cm)},
    y={(0cm,-1cm)},
    z={(0.35cm,0.25cm)},
    xshift=10*\cellsize cm, yshift=-5.5*\cellsize cm,
    scale=0.8
]


    \draw[fill=white, draw=black] (-0.5*\cellsize,0.5*\cellsize,20*\cellsize) rectangle ++(3.5*\cellsize,-3*\cellsize,0);
    
    \draw[fill=Cell2h4, draw=black, Hatch] (0*\cellsize,0,24*\cellsize) rectangle ++(\cellsize,-\cellsize,0);
    \draw[fill=Cell2h5, draw=black, Hatch] (0*\cellsize,0,23*\cellsize) rectangle ++(\cellsize,-\cellsize,0);
    \draw[fill=Cell2h3, draw=black] (0*\cellsize,0,22*\cellsize) rectangle ++(\cellsize,-\cellsize,0);
    \draw[fill=Cell2h2, draw=black] (0*\cellsize,0,21*\cellsize) rectangle ++(\cellsize,-\cellsize,0);
    \draw[fill=Cell2h1, draw=black] (0*\cellsize,0,20) rectangle ++(\cellsize,-\cellsize,0);
    

    \draw[fill=white, draw=black] (-0.5*\cellsize,0.5*\cellsize,15*\cellsize) rectangle ++(3.5*\cellsize,-3*\cellsize,0);
    
    \draw[fill=Cell3h4, draw=black, Hatch] (0*\cellsize,0,19*\cellsize) rectangle ++(\cellsize,-\cellsize,0);
    \draw[fill=Cell3h5, draw=black, Hatch] (0*\cellsize,0,18*\cellsize) rectangle ++(\cellsize,-\cellsize,0);
    \draw[fill=Cell3h3, draw=black] (0*\cellsize,0,17*\cellsize) rectangle ++(\cellsize,-\cellsize,0);
    \draw[fill=Cell3h2, draw=black] (0*\cellsize,0,16*\cellsize) rectangle ++(\cellsize,-\cellsize,0);
    \draw[fill=Cell3h1, draw=black] (0*\cellsize,0,15) rectangle ++(\cellsize,-\cellsize,0);


    \draw[fill=white, draw=black] (-0.5*\cellsize,0.5*\cellsize,10*\cellsize) rectangle ++(3.5*\cellsize,-3*\cellsize,0);
    
    \draw[fill=Cell5h4, draw=black, Hatch] (0*\cellsize,0,14*\cellsize) rectangle ++(\cellsize,-\cellsize,0);
    \draw[fill=Cell5h5, draw=black, Hatch] (0*\cellsize,0,13*\cellsize) rectangle ++(\cellsize,-\cellsize,0);
    \draw[fill=Cell5h3, draw=black] (0*\cellsize,0,12*\cellsize) rectangle ++(\cellsize,-\cellsize,0);
    \draw[fill=Cell5h2, draw=black] (0*\cellsize,0,11*\cellsize) rectangle ++(\cellsize,-\cellsize,0);
    \draw[fill=Cell5h1, draw=black] (0*\cellsize,0,10) rectangle ++(\cellsize,-\cellsize,0);


    \draw[fill=white, draw=black] (-0.5*\cellsize,0.5*\cellsize,5*\cellsize) rectangle ++(3.5*\cellsize,-3*\cellsize,0);
    
    \draw[fill=Cell4h4, draw=black, Hatch] (0*\cellsize,0,9*\cellsize) rectangle ++(\cellsize,-\cellsize,0);
    \draw[fill=Cell4h5, draw=black, Hatch] (0*\cellsize,0,8*\cellsize) rectangle ++(\cellsize,-\cellsize,0);
    \draw[fill=Cell4h3, draw=black] (0*\cellsize,0,7*\cellsize) rectangle ++(\cellsize,-\cellsize,0);
    \draw[fill=Cell4h2, draw=black] (0*\cellsize,0,6*\cellsize) rectangle ++(\cellsize,-\cellsize,0);
    \draw[fill=Cell4h1, draw=black] (0*\cellsize,0,5) rectangle ++(\cellsize,-\cellsize,0);


    \draw[fill=white, draw=black] (-0.5*\cellsize,0.5*\cellsize,0*\cellsize) rectangle ++(3.5*\cellsize,-3*\cellsize,0);
    
    \draw[fill=Cell1h4, draw=black, Hatch] (0*\cellsize,0,4*\cellsize) rectangle ++(\cellsize,-\cellsize,0);
    \draw[fill=Cell1h5, draw=black, Hatch] (0*\cellsize,0,3*\cellsize) rectangle ++(\cellsize,-\cellsize,0);
    \draw[fill=Cell1h3, draw=black] (0*\cellsize,0,2*\cellsize) rectangle ++(\cellsize,-\cellsize,0);
    \draw[fill=Cell1h2, draw=black] (0*\cellsize,0,1*\cellsize) rectangle ++(\cellsize,-\cellsize,0);
    \draw[fill=Cell1h1, draw=black] (0*\cellsize,0,0) rectangle ++(\cellsize,-\cellsize,0);

    \draw[draw=black, <->, >=stealth] (-0.75,-3.25,0) -- ++(0,0,20*\cellsize) node[midway, above, rotate=37, scale=0.75] {\revI{$2N^{[2]}-1$}};

    \foreach \x in {1,2,3}
        \draw (-0.65,-3.15,5*\x*\cellsize) -- (-0.85,-3.35,5*\x*\cellsize);

\end{scope}


\end{tikzpicture}
    \caption{\rev{Visualization of conversion from 2D and 4D data structure \revI{between \figref{fig:ToepStruct}b and \figref{fig:ToepStruct}c for a $3\multX3$ array, i.e., $\mli{N}{2}=\mli{N}{1}=3$}. (a) One set of blocks on level 1. (b) Blocks on level 1 stacked along the third dimension. (c) Level 1 structures stacked along the fourth dimension.}}
    \label{fig:4Dillustration}
\end{figure}

\rev{To visualize this structure, \revI{see} \figref{fig:4Dillustration}. Start with one set of data on level 1 \revI{from \figref{fig:4Dillustration}b, i.e., one colored set}. Each block on the lowest level is kept as a 2D matrix along dimensions 1 and 2. The index on level 1 is assigned to dimension 3, resulting in the set having the structure in \figref{fig:4Dillustration}b. This is done for each set of data on level 1, and these are then stacked along dimension 4 as illustrated in \figref{fig:4Dillustration}c.} Then the MLFFT \rev{can instead be performed by} applying two FFTs along dimensions 3 and 4. \rev{Instead of allocating temporary storage and indexing between several rows in memory, this application of MLFFT \revI{is aligned with the FFT implementation and consequentially} much better optimized.} \rev{By reshaping the 4D structure to a vector \revI{similar to \cite{hultin2025solver},} it can still be passed to \revI{GMRES and PCG implemetations} that require vectors as an input.} The resulting vectorization is \revII{a permutation to the result} in \cite{hultin2025solver} \revI{as the 4D conversion reorders the data}.

\subsection{\revIV{Solver setup}}
\label{sec:solversetup}
\revIV{In this publication, two new solvers are introduced. These solvers are denoted as "new M1", as it builds on the GMRES-based solver "M1" in \cite{hultin2025solver}, and "PCG" that uses the same improvements as new M1, but in a PCG-based solver. In \figref{fig:flownew}, the changes to the new solvers are shown, in contrast to the previous implementation of M1 in \figref{fig:flowold}.}

\begin{figure}
    \centering
    \subfloat[Flowchart of new solvers "new M1" and "PCG"]{
    \begin{tikzpicture}[ 
    node distance=4mm, 
    >=Latex, 
    block/.style={ 
        draw, 
        rounded corners, 
        align=center, 
        font=\footnotesize, 
        minimum width=3.2cm, 
        minimum height=6mm, fill=gray!10 
        } 
    ] 
    
    \node[block] (z) {Compute elements of $Z$ \cite{akerstedt2024partitioning} \\Specify $V$ in $ZI=V$}; 
    
    \node[block, right=of z] (prec) {Compute preconditioner $P$ from \\$Z$ as specified in \secref{sec:precondcalc}}; 

    \node[block, below=of prec] (chol) {Cholesky factorize $P=LL^H$};
    
    \node[block, left=of chol] (mlfft) {Precompute MLFFT of $Z_A$ with \\improvements from \secref{sec:dataStruct}}; 

    \node[block, below=of mlfft] (funcZ) {Define function $u= Zv$ as \\in \cite{hultin2025solver} with improved MLFFT \\from \secref{sec:dataStruct}}; 

    \node[block, right=of funcZ] (funcZH) {Define function $u= Z^Hv$ using \\property from \secref{sec:hermprop}}; 
    
    \node[block, below=of funcZH] (funcs) {Define functions\\ $u=Z^HZv$ \\$u=L^{-1}v$ \\$u=L^{-H}v$}; 
    
    \node[block, left=of funcs] (solver) {GMRES / PCG\\  input: $V$ \& function handles}; 

    \node[block, below=of solver] (solution) {Solution $I$}; 
    
    \draw[->] (z) -- (prec); 
    
    \draw[->] (prec) -- (chol); 

    \draw[->] (chol) -- (mlfft); 
    
    \draw[->] (mlfft) -- (funcZ); 

    \draw[->] (funcZ) -- (funcZH); 

    \draw[->] (funcZH) -- (funcs);
    
    \draw[->] (funcs) -- (solver); 

    \draw[->] (solver) -- (solution);
    
    
\end{tikzpicture}
    \label{fig:flownew}}
    
    \subfloat[Flowchart of old solver "M1"]{
    \begin{tikzpicture}[ 
    node distance=4mm, 
    >=Latex, 
    block/.style={ 
        draw, 
        rounded corners, 
        align=center, 
        font=\footnotesize, 
        minimum width=3.2cm, 
        minimum height=6mm, fill=gray!10 
        } 
    ] 
    
    \node[block] (z) {Compute elements of $Z$ \cite{akerstedt2024partitioning} \\Specify $V$ in $ZI=V$}; 
    
    \node[block, right=of z] (prec) {Extract preconditioner $P_K$ \\from $Z$ as in \cite{hultin2025solver}}; 

    \node[block, below=of prec] (LU) {LU-factorize $P=LU$ \\and define functions\\$u=L^{-1}v$, $u=U^{-1}v$};
    
    \node[block, left=of LU] (mlfft) {Precompute MLFFT of $Z_A$, define \\function $u= Zv$ as in \cite{hultin2025solver}}; 
    
    \node[block, below=of mlfft] (solver) {GMRES\\  input: $V$ \& function handles}; 

    \node[block, right=of solver] (solution) {Solution $I$}; 
    
    \draw[->] (z) -- (prec); 
    
    \draw[->] (prec) -- (LU); 

    \draw[->] (LU) -- (mlfft); 
    
    \draw[->] (mlfft) -- (solver); 

    \draw[->] (solver) -- (solution);
    
    
\end{tikzpicture}
    \label{fig:flowold}}
    \caption{\revIV{Flowchart of how solvers progress from $Z$ given by \cite{akerstedt2024partitioning} and $V$, to a solution $I$.}}
    \label{fig:flowchart}
\end{figure}

\revIV{As in the previous implementation, the methods start with $Z$ from \eqref{eq:Zblock} as defined in \cite{akerstedt2024partitioning}, and a matrix $V$, where the columns contain excitation vectors as introduced in \secref{sec:hermprop}. When defining the preconditioner, M1 extracted the block-diagonal of $Z$ to create $P_K$, whereas the new solvers need to calculate the block-diagonal of $Z^HZ$ without explicitly forming the matrix as described in \secref{sec:precond} and \secref{sec:precondcalc}. As $P_K$ is non-Hermitian, it cannot be Cholesky factorized and is therefore factorized as two matrices $L$ and $U$, unlike the new preconditioner which can be represented by a matrix $L$ alone. Note that, as each element excitation is solved separately, an arbitrary array excitation can be found by combining the element excitations in post processing. Furthermore, this means that the convergence of the solver is independent of the chosen array excitation.}

\revIV{For both solvers, the MLFFT of $Z_A$ is precomputed, but on the new data format of \secref{sec:dataStruct} in the new solvers. This is then used to define a multiplication of $Z$, which is also improved by the new data structure. In addition, the new solver requires a multiplication by $Z^H$ as well, which can be carried out without saving a new data set as described in \secref{sec:hermprop}. The functions for multiplying by $Z$ or $Z^HZ$, together with functions for applying the inverse of the preconditioner, are then passed to either GMRES or PCG depending on the solver. Here the inverses are applied using a block-wise use of the function \emph{mldivide} in MATLAB, and GMRES and PCG are implemented using the built in functions \emph{gmres} and \emph{pcg} \cite{MATLAB_R2024a}}.

\revIV{In the MATLAB documentation for their implementation of PCG, it is specified that the function effectively solves the problem as specified at the end of \secref{sec:precondcalc} \cite{MATLAB_R2024a}, although the principal square root of $P$ is never explicitly calculated in the method implementation.}

\revIV{The preconditioner $P$ has the same size as the preconditioner $P_K$ in \cite{hultin2025solver}, and is therefore not expected to be a leading factor in the scaling of the computational complexity. While the new method uses two MLFFT accelerated multiplications instead of one, this still scales as $\mathcal{O}(N^2\log (N))$ for an $N$ element array with $N$ ports, or $\mathcal{O}(N\log (N))$ for a single excitation. This multiplication will still be the leading term as in \cite{hultin2025solver}, and it is therefore expected to still scale similarly as, e.g., other Toeplitz solvers \cite{brandt2024extended, cavillot2019efficient}, but the improvements lead to a smaller constant factor in front of the scaling.}

\subsection{Antennas}
\label{sec:antennas}
\begin{figure}[t]
\centering
\subfloat[]{
    \resizebox{0.35\columnwidth}{!}{%
        \begin{tikzpicture}[scale=3, thick]
        
        \draw[dotted] (-0.55,-0.55) -- (-0.55,0.55) -- (0.55,0.55) -- (0.55,-0.55) -- (-0.55,-0.55);
    
        \draw[thick] (-0.5,-0.25) -- (-0.5,0.25) -- (0,0.01) -- (0.5,0.25) -- (0.5,-0.25) -- (0,-0.01) -- (-0.5,-0.25);
        
        \draw[Set1-A, ultra thick] (0,-0.01) -- (0,0.01);
    
        \draw[<->] (-0.5,0.4) -- node[below] {$0.5\lambda$}(0.5,0.4);
        \draw[<->] (-0.55,-0.5) -- node[above] {$0.55\lambda$}(0.55,-0.5);
        \draw[<->] (0.55,0.25) -- node[above, rotate=270] {$0.25\lambda$}(0.55,-0.25);
        \draw[<->] (0.7,-0.55) -- node[above, rotate=270] {$0.55\lambda$}(0.7,0.55);

        \end{tikzpicture}
    }\label{fig:bowtie_geo}
    
}
\subfloat[]{
    \includegraphics[]{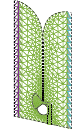}
    \label{fig:viv_geo}
    
}
\subfloat[]{
    \includegraphics[]{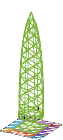}
    \label{fig:bor_geo}
    
}

\caption{Antenna unit cell geometries. (a) Bowtie unit cell from above, \rev{defined at \SI{100}{\mega\hertz}}. The bowtie is placed $0.25\lambda$ over a ground plane. (b) Vivaldi elements from \cite{akerstedt2024partitioning} with a \revII{center} frequency of \SI{8}{\giga\hertz}. (c) BoR element from \cite{akerstedt2024partitioning} with an upper frequency of \SI{18}{\giga\hertz}. \revI{Both (b) and (c) have $\lambda/2$ spacing at the upper frequency.}} 
\label{fig:antennas}
\end{figure}

Three different array antenna examples will be used to generate the results: a bowtie, \revI{a Vivaldi, and a} BoR array. \revI{Their elements are illustrated in \figref{fig:antennas}.} The bowtie elements \revI{in \figref{fig:bowtie_geo}} are defined at a frequency of \rev{\SI{100}{\mega\hertz}} and lie $0.25\lambda$ above a finite ground plane. The Vivaldi element is designed for \SI{8}{\giga\hertz} and the BoR for \SI{18}{\giga\hertz}\rev{, both are described more thoroughly} in \cite{akerstedt2024partitioning}. Solutions for the bowtie and Vivaldi elements are found using the preconditioner in \secref{sec:precond} and PCG. The BoR uses the data set from \cite{akerstedt2024partitioning} as a reference solution, which has been solved using a direct method.

\section{Results}
\label{sec:results}
\subsection{\rev{Simulation improvements}}
\revII{The preconditioning strategy of \secref{sec:precond} and the improved data structure of \secref{sec:dataStruct}, enables efficient simulation of large array antennas. The performance is first evaluated using the bowtie array shown in \figref{fig:bowtie_geo}. This element was previously investigated in \cite{hultin2025solver}, where arrays based on this element resulted in the slowest convergence of the GMRES-based solver M1.}

\begin{figure}[t]
    \centering
%
%
\colorlet{mycolor1}{black}%
\colorlet{mycolor2}{Set1-A}%
\colorlet{mycolor3}{Set1-B}%
\colorlet{mycolor4}{Set1-C}%
\colorlet{mycolor5}{Set1-D}%
\colorlet{mycolor6}{Set1-E}%
\colorlet{mycoloryellow}{Set1-F}%
\colorlet{mycolor7}{Set1-G}%
\begin{tikzpicture}

\begin{axis}[%
width=\columnwidth,
height=5.8cm,
xmode=log,
xmode=log,
xmin=16,
xmax=1024,
xminorticks=true,
xlabel style={font=\color{white!15!black}},
xlabel={$\text{Number of array elements}$},
xmajorgrids,
xminorgrids,
ymajorgrids,
ymode=log,
ymin=0.1,
ymax=100000,
ytick={1, 10, 100, 1000, 10000, 100000},
yminorticks=true,
ylabel style={font=\color{white!15!black}},
ylabel={time (s)},
axis background/.style={fill=white},
legend style={at={(1,0)}, anchor=south east, legend cell align=left, align=left, draw=white!15!black, legend columns=3}
]

\addplot [color=mycolor2, very thick, mark=x]
  table[row sep=crcr]{%
16	3.3422502\\
36	21.534735\\
64	104.12991\\
196	2102.8708813\\
};
\addlegendentry{mld}

\addplot [color=mycolor3, very thick, mark=o]
  table[row sep=crcr]{%
16	29.909003\\
36	278.3542488\\
64.0000000000001	1540.4187475\\
196	50008.7018715\\
};
\addlegendentry{G}


\addplot [color=mycolor5,very thick, mark=|]
  table[row sep=crcr]{%
16	10.8117086\\
36	76.3513432999999\\
64.0000000000001	372.8108777\\
196	7398.94328370001\\
400	50147.5878972\\
};
\addlegendentry{M1}

\addplot [color=mycolor6,very thick, mark=|]
  table[row sep=crcr]{%
16	6.5 \\
64	60\\
256	1591\\
576	11406\\
1024 38000 \\
};
\addlegendentry{PCG}

\addplot [color=mycolor5,very thick, mark=*]
  table[row sep=crcr]{%
16	3.367\\
64	46.26\\
256	1212\\
576	9063\\
961 42496\\
};
\addlegendentry{new M1}




\addplot [color=mycolor7,very thick, mark=*]
  table[row sep=crcr]{%
16	0.5\\
64	3\\
256	22\\
576	82\\
961 201\\
};
\addlegendentry{\revIV{$P$}}

\end{axis}

\end{tikzpicture}%
    \caption{\revI{Solution} time for \revI{bowtie arrays} over number of array elements at \SI{150}{\mega\hertz} with \num{E-3} tolerance for iterative solvers. mld: Matlab linear equation system solver, G: GMRES, M1: old iterative solver, PCG: PCG \revI{with new preconditioning scheme}, new M1: M1 with new \revI{preconditioning scheme}\revIV{, $P$: time to construct and factorize preconditioner $P$}.}
    \label{fig:lightBTscaling}
\end{figure}
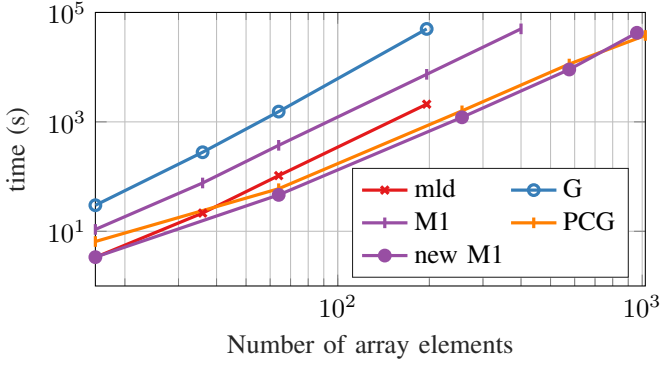

One way to incorporate the improvements is to simply use them in GMRES, i.e. a similar solver to M1 in \cite{hultin2025solver} but with \revII{the new preconditioning strategy in \secref{sec:precond} and the new data structure in \secref{sec:dataStruct}}. \revIV{As introduced in \secref{sec:solversetup}, this solver is} denoted "new M1". But as discussed in \secref{sec:PCG}, the preconditioning scheme makes the problem suitable to solve using PCG instead of GMRES. Unlike GMRES, PCG has a low memory usage that is independent of the number of iterations. These two ways of solving the problem are now the fastest, \revI{reducing the solution time by at least a factor 8 compared to M1 for arrays with more than 64 elements} as seen in \figref{fig:lightBTscaling}. The PCG solver shows a more stable scaling compared to the new M1 in addition to \revII{using less memory} \revIV{compared to M1 as its memory usage does not grow with the number of iterations. As stated in \secref{sec:precondcalc}, constructing $P$ only takes a fraction of the time compared to solving the problem.} 

In part, the improvement in solver speed is from the improved preconditioning. But it should be noted that preconditioning methods can affect the numerical accuracy of the solvers. As an example, GMRES with the preconditioner $P_K$ from \cite{hultin2025solver}, a preconditioner consisting of the block diagonal elements on the lowest level in the MoM-matrix $Z$, gives an relative rms error (RRMSE) in the passive impedance matrix $\revI{Z_p}$ on a similar scale as the convergence criteria \num[]{e-3} as shown in Table~\ref{tab:errors}. \rev{The error is defined as}
\begin{equation}
    \rev{RRMSE = \sqrt{\frac{1}{N_P}\sum_{n=1}^{N_P}\frac{|[\Zp{\text{solver}}]_n-[\Zp{\text{direct}}]_n|}{[\Zp{\text{direct}}]_n}},}
\end{equation}
\rev{where $N_P$ is the number of ports, $[\Zp{\text{solver}}]_n$ is the port impedance of port $n$ using an iterative solver and $[\Zp{\text{direct}}]_n$ is the port impedance of port $n$ using a direct solver}. However, if the preconditioning scheme $P^{-1}Z^H$ introduced in \secref{sec:precond} is used with GMRES, i.e. new M1, \rev{RRMSE} grows by a factor $10$ \rev{compared to $P_K^{-1}Z$ as seen in Table~\ref{tab:errors}, even though the convergence criteria is kept the same}. But if PCG is used instead, which is well suited for Hermitian matrices and which applies the preconditioner $P$ in a different manner, the error stays on a similar level as for GMRES with $P_K$. Note that this is the \revII{RRMSE} for the whole matrix $\revI{Z_p}$, the \revII{relative} error for the center element that is mainly considered here is \num[]{1.7e-4}. \revII{In addition to improved stability and lower memory requirement, PCG appears to preserve numerical accuracy better than new M1, while remaining faster than the old M1.}

\begin{table}[t]
    \centering
    \caption{Relative rms error in $\revI{Z_p}$ compared to direct solution \revIV{for a $6\multX 6$ bowtie}}
    \begin{tabular}{llll}
        \hline
         \rev{\textbf{Solver}} & \rev{\textbf{Matrix}} & \textbf{\rev{RRMSE}} & \revIV{\textbf{Condition number}} \\
         \hline
         \revIV{M1} & $P_K^{-1}Z$ & \num[]{3.8e-3} & \revIV{\num{5.4e+2}}\\
         \revIV{new M1} & $P^{-1}Z^HZ$ & \num[]{4.6e-2} & \revIV{\num{1.2e+3}} \\
         PCG & $P^{-\frac{1}{2}}Z^HZP^{-\frac{H}{2}}$ & \num[]{4.5e-3} & \revIV{\num{1.2e+3}} \\
         \hline
    \end{tabular}
    \label{tab:errors}
\end{table}

\begin{figure}[t]
    \centering
    \includegraphics[width= 0.9\columnwidth]{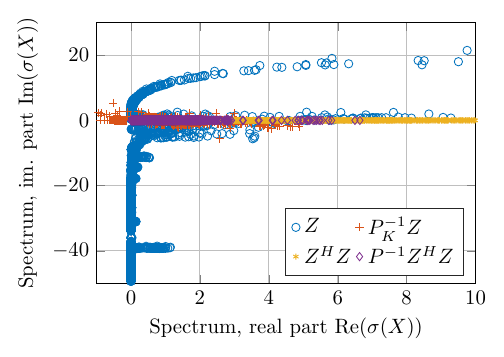}
    \caption{\revI{Spectrum of complex eigenvalues $\sigma(X)$} of the MoM-matrix with and without preconditioners \rev{for a $6\multX 6$ bowtie array at \SI{100}{\mega\hertz}}. \revI{Here, $X$ is the legend entry.} The preconditioner $P_K$ is the block diagonal on the lowest level as in \cite{hultin2025solver}. The x-axis is cut of at 10, \revI{even though $Z^HZ$ has a dense spectrum with real eigenvalues as large as 2500. All other spectrum are completely contained within the shown plane. Note that $P^{-1}Z^HZ$ has the same spectrum as $P^{-\frac{1}{2}}Z^HZP^{-\frac{H}{2}}$}.}
    \label{fig:spectrum}
\end{figure}

Some insight to why the preconditioner affects the error can be found in the spectrum of the matrix that are solved. For the MoM-matrix $Z$, the eigenvalues show some structured grouping, but they are spread out and have relatively large complex components as shown in \figref{fig:spectrum} \revIV{with a condition number of \num{1.7E+03}}. Both of these properties can slow down iterative solvers \cite{carson2024towards}. Applying $P_K$ puts the eigenvalues much closer to the real axis and in a narrower range \revIV{and reduces the condition number to \num{5.4e+2}}. With the preconditioner $P^{-1}Z^HZ$, which has the same spectrum\revI{\footnote{\revI{See \secref{sec:precondcalc}.}}} as $P^{-\frac{1}{2}}Z^HZP^{-\frac{H}{2}}$, the eigenvalues lie on the real axis in a narrow spectrum, and in groups. All of these three properties are favorable to solve the problem fast. \revIV{The condition number shrinks to \num{1.2e+3}, i.e., not as much as for $P_K$. But as evident from the results and pointed out in \cite{carson2024towards}, the condition number alone does not determine the performance of a preconditioner.} If $Z^HZ$ alone is considered, the eigenvalues \revII{have} a very broad and dense spectrum. Note that the x-axis is cut of at 10, the eigenvalues of $Z^HZ$ continue up to 2500, as the eigenvalues have the squared amplitude of those of $Z$. This in turn gives the matrix a \revIV{conditioning number of \num{1.3E+4}}, that can impact numerical accuracy as seen for GMRES \rev{ when solving $P^{-1}Z^HZ$} in Table~\ref{tab:errors}, even though the total preconditioning scheme results in a favorable spectrum. But using the scheme correctly, as in PCG, it becomes a powerful tool. 

\subsection{\rev{Asymptotic properties of array antennas}}
With a fast and efficient full-wave \rev{PCG} solver, large arrays can be solved without making macroscopic approximations on element behavior\rev{, compared to, e.g., unit cell methods where elements are approximated to behave identically in several aspects}. But how important are full-wave solvers for elements closer to the center of the array as the array grows larger? As discussed in \secref{sec:introduction} there are several methods based on making macroscopic approximations and assumptions for sufficiently large arrays. One popular thumb rule for what "sufficiently large" is states that a $10\multX 10$ array is large enough to start making approximations on element behavior. This rule can be generalized to $5\lambda \multX 5 \lambda$ \cite{HolterH.2002Otsr}.

\begin{figure}[t]
    \centering
%
%
\colorlet{mycolorb}{black}%
\colorlet{mycolor1}{Set1-B}%
\colorlet{mycolor2}{Set1-A}%
\colorlet{mycolor3}{Set1-C}%
\colorlet{mycolor5}{Set1-D}%
\colorlet{mycolor6}{Set1-E}%
\colorlet{mycolor4}{Set1-F}%
\colorlet{mycolor7}{Set1-G}%
\begin{tikzpicture}

\begin{axis}[%
width=\columnwidth,
height=5.5cm,
xmin=2,
xmax=14,
xlabel style={font=\color{white!15!black}},
xlabel={\revII{Electrical} array size ($\lambda$)},
ymin=0.8,
ymax=1.19,
ylabel={\revII{Normalized active impedance}},
 y tick label style={
        /pgf/number format/.cd,
            fixed,
            fixed zerofill,
            precision=2,
        /tikz/.cd
    },
axis background/.style={fill=white},
title style={font=\bfseries},
minor x tick num=4,
xmajorgrids,
xminorgrids,
ymajorgrids,
legend style={at={(0.97,0.03)}, anchor=south east, legend cell align=left, align=left, draw=white!15!black, legend columns=2},
]
\addplot [color=mycolor1, very thick]
  table[row sep=crcr]{%
2.1978021978022	0.944036023022056\\
2.74725274725275	1.07865137439286\\
3.2967032967033	1.01398528952846\\
3.84615384615385	0.951408245968841\\
4.3956043956044	0.977523174453635\\
4.94505494505495	1.01222372385229\\
5.49450549450549	0.995772189876888\\
6.04395604395604	0.982288176352087\\
6.59340659340659	0.986648115310823\\
7.14285714285714	0.987983089424194\\
7.69230769230769	0.99144667006353\\
8.24175824175824	0.993039388274669\\
8.79120879120879	0.988852069742229\\
9.34065934065934	0.985606539495155\\
9.89010989010989	0.989787390638552\\
10.4395604395604	0.995286982675674\\
11.5384615384615	0.984328940551604\\
12.0879120879121	0.989345285600683\\
12.6373626373626	0.993353122267999\\
13.1868131868132	0.990209927027333\\
};

\addlegendentry{BT \SI{100}{\mega\hertz}}

\addplot [color=mycolor5, very thick]
  table[row sep=crcr]{%
0.733333333333333	1.58202147220065\\
1.46666666666667	1.31408643306812\\
2.20000000000000	0.994147884929093\\
2.93333333333333	0.939928660964843\\
3.66666666666667	0.906762413715160\\
4.40000000000000	0.932178313016651\\
5.13333333333333	0.980670547628479\\
5.86666666666667	0.987519288661098\\
6.60000000000000	1.00737226046113\\
7.33333333333333	0.987371591495932\\
8.06666666666667	0.974362558354074\\
8.80000000000000	0.966277302632651\\
9.53333333333333	0.963449262822178\\
10.2666666666667	0.969802301573495\\
11	0.980108964541103\\
11.7333333333333	0.983010925156863\\
12.4666666666667	0.987193945745940\\
13.2000000000000	0.981601628714566\\
13.9333333333333	0.976523737478065\\
14.6666666666667	0.975332272334412\\
15.4000000000000	0.974678414233465\\
16.1333333333333	0.980259054213811\\
16.8666666666667	0.988698034144706\\
17.6000000000000	0.993282528389579\\
18.3333333333333	0.986223814508623\\
19.0666666666667	0.979557837190456\\
19.8000000000000	0.985676909749558\\
20.5333333333333	0.986708607320321\\
21.2666666666667	0.990454008401259\\
22	0.995933914953467\\
22.7333333333333	1\\
};
\addlegendentry{BT \SI{150}{\mega\hertz}}

\addplot [color=mycolor2, very thick]
  table[row sep=crcr]{%
2.13333333333333	1.11864683723422\\
2.66666666666667	1.1748084580072\\
3.2	1.01781117677282\\
3.73333333333333	0.881091155053664\\
4.26666666666667	0.928475803732483\\
4.8	0.961387921851658\\
5.33333333333333	1.02063536943732\\
5.86666666666667	1.04689916418527\\
6.4	0.962304518414135\\
};
\addlegendentry{Viv. \SI{8}{\giga\hertz}}

\addplot [color=mycolor3, very thick]
  table[row sep=crcr]{%
3.2	0.727515126335168\\
4	0.961191118492515\\
4.8	1.04986324066111\\
5.6	1.14007456609806\\
6.4	1.03163247945491\\
7.2	1.02645009660944\\
8	1.07775015696408\\
8.8	1.09725839053108\\
9.6	0.927684988047261\\
};
\addlegendentry{Viv. \SI{12}{\giga\hertz}}

\addplot [color=white!15!black, very thick, forget plot]
  table[row sep=crcr]{%
5	0\\
5	2\\
};

\end{axis}

\end{tikzpicture}%
    \caption{\revII{Normalized} active impedance \revV{$|Z_{a,c}/Z_{a,0}|$} of the center element for boresight excitation for different elements and frequencies over varying array sizes. Rule of thumb $5\lambda\multX 5\lambda$ \revV{is highlighted as a thick black line}. \revI{The active impedance is normalized to the average of the 10 largest array sizes \revV{$Z_{a,0}$.}}}
    \label{fig:normImp}
\end{figure}
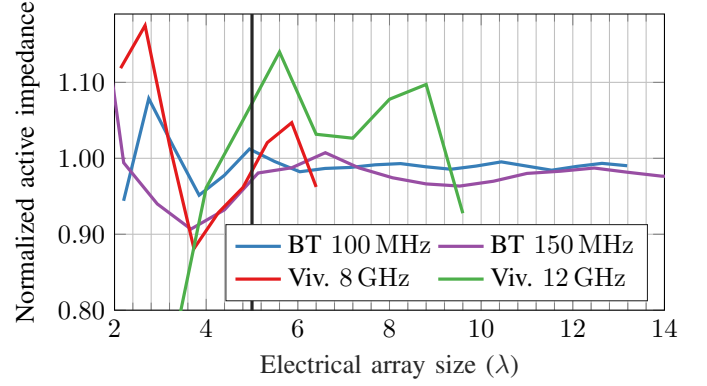

To see how consistent these thumb rules are for \rev{active} impedance, a simple test is to see how the active impedance behaves for the center element as the array size varies. This is shown in \figref{fig:normImp} for the bowtie and Vivaldi arrays at two different frequencies each. \revI{The higher frequencies are intended to gauge how well the $5\lambda$ rule works for a spacing other than $\lambda/2$, while still being within the \revII{working} band of the antenna.} 

For the bowtie element, the  impedance converges rather quickly, with a  relatively flat error for sizes larger than $5\lambda$. \revI{The active impedance is normalized to the average \revV{$\Za{0}$} of the 10 largest array sizes.} The Vivaldi shows more variations, and it is overall difficult to see any type of consistent decay over size. These \revV{variations} can lead to incorrect assumptions, and it would therefore be advantageous to predict an estimate of the deviation before making approximations or assumptions.

As discussed in \secref{sec:introduction} and described in \secref{sec:predict} the passive scattering parameter can be used for this purpose. \revI{The} parameter $S_{mn}$  between elements $m$ and $n$ can be extracted or predicted in several ways. A infinite array is considered, with a single "center" element, $m=0$, excited. \revI{Let $n$ be another element in the same row} with $n=1$ being the neighboring element, $n=2$ being one element further away, and so on. \revI{Then, $\elDist=n$.} For infinite arrays, it has been shown that the coupling coefficient decreases monotonically with a slope of $\elDist^{-2}$ for $\elDist\gtrapprox 5$ \cite{amitay1972theory, galindo1968asymptotic}. It is also stated that "the asymptotic dependence of the coupling coefficients seems to be universal for all phased arrays" \cite{amitay1972theory}. This type of behavior would strengthen the $10\multX 10$ array theory, or the more general $5\lambda\multX 5\lambda$ spacing for arrays with with spacings other than $\lambda/2$ \cite{HolterH.2002Otsr}. While this does not give a clear answer to how large the deviations are expected to be at a given size, it is a step on the way to a predictor of how the deviation depends on array size.

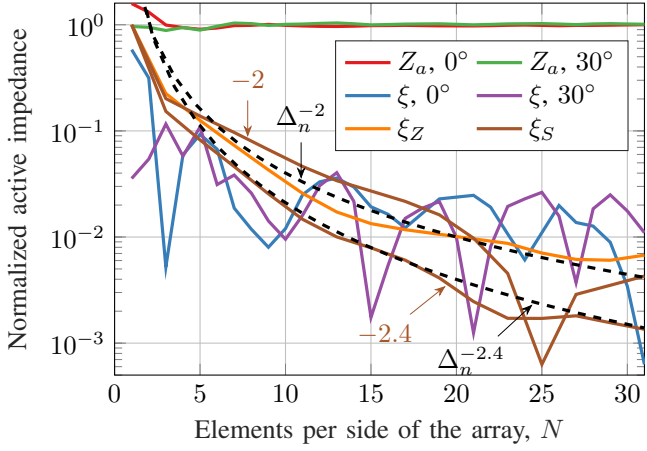
\begin{figure}[t]
    \centering
%
%
\colorlet{mycolor1}{Set1-A}%
\colorlet{mycolor2}{Set1-B}%
\colorlet{mycolor3}{Set1-C}%
\colorlet{mycolor4}{Set1-D}%
\colorlet{mycolor5}{Set1-E}%
\colorlet{mycoloryellow}{Set1-F}%
\colorlet{mycolor6}{Set1-G}%
\begin{tikzpicture}

\begin{axis}[%
width=0.97\columnwidth,
height=6.5cm,
xmin=0,
xmax=31,
xlabel style={font=\color{white!15!black}},
xlabel={\revII{Elements per side of the array,} $N$},
ymin=0.0005,
ymax=1.6,
ymode=log,
ylabel style={font=\color{white!15!black}},
ylabel={Normalized \revII{active} impedance},
axis background/.style={fill=white},
title style={font=\bfseries},
xmajorgrids,
ymajorgrids,
domain=1:31,
legend style={at={(0.97,0.9)}, anchor=north east, legend cell align=left, align=left, draw=white!15!black, legend columns=2}
]
\addplot [color=mycolor1, very thick]
  table[row sep=crcr]{%
1	1.58301907699641\\
2	1.31491508106735\\
3	0.994774783308873\\
4	0.940521369316993\\
5	0.907334207807992\\
6	0.932766134087295\\
7	0.98128894735223\\
8	0.988142007123262\\
9	1.00800749798212\\
10	0.987994216821939\\
11	0.974976980331338\\
12	0.966886626139367\\
13	0.964056802999067\\
14	0.97041384790456\\
15	0.980727010136928\\
16	0.983630800696137\\
17	0.987816459050549\\
19	0.97713952227037\\
21	0.975293035385864\\
22	0.980877194454482\\
23	0.989321495910399\\
24	0.993908881085343\\
26	0.980175535251625\\
27	0.986298466428476\\
28	0.987330814576026\\
29	0.9910785774644\\
30	0.996561939583522\\
31	1.00063058865717\\
};
\addlegendentry{\revI{$Z_a$, \ang{0}}}

\addplot [color=mycolor3, very thick]
  table[row sep=crcr]{%
1	0.964373841874339\\
2	0.94577119238213\\
3	0.883932591975195\\
4	0.942074271552161\\
5	0.890995024360507\\
6	0.968793055805687\\
7	1.03835665169555\\
8	1.02595718566938\\
9	0.985821018068336\\
10	1.00952606503598\\
11	1.01603379548426\\
12	1.03047138686516\\
13	1.04040160533177\\
14	1.02164396441779\\
15	0.998250584721272\\
16	1.00540517992051\\
17	1.01496679838718\\
18	1.01867875161449\\
19	1.02159451023494\\
20	1.00992141837536\\
21	0.998747641787493\\
22	1.00813876218385\\
23	1.01942171494495\\
24	1.02288138170092\\
25	1.02636481790508\\
26	1.01578445745251\\
27	1.00372305535934\\
28	1.01845124281725\\
29	1.02496243467936\\
30	1.01757385132284\\
31	1.01098174607441\\
};
\addlegendentry{\revI{$Z_a$, \ang{30}}}

\addplot [color=mycolor2, very thick]
  table[row sep=crcr]{%
1	0.583019076996411\\
2	0.314915081067351\\
3	0.00522521669112663\\
4	0.0594786306830066\\
5	0.0926657921920082\\
6	0.067233865912705\\
7	0.0187110526477703\\
8	0.0118579928767417\\
9	0.00800749798211697\\
10	0.0120057831780649\\
11	0.0250230196686623\\
12	0.0331133738606333\\
13	0.0359431970009325\\
14	0.0295861520954404\\
15	0.0192729898630724\\
16	0.0163691993038633\\
17	0.0121835409494508\\
19	0.0228604777296297\\
21	0.0247069646141362\\
22	0.0191228055455177\\
23	0.0106785040896007\\
24	0.0060911189146573\\
26	0.0198244647483747\\
27	0.013701533571524\\
28	0.0126691854239738\\
29	0.00892142253560024\\
30	0.00343806041647809\\
31	0.000630588657166697\\
};
\addlegendentry{\revI{$\xi$, \ang{0}}}

\addplot [color=mycolor4, very thick]
  table[row sep=crcr]{%
1	0.0356261581256606\\
2	0.0542288076178694\\
3	0.116067408024805\\
4	0.0579257284478394\\
5	0.109004975639493\\
6	0.0312069441943132\\
7	0.0383566516955507\\
8	0.025957185669383\\
9	0.0141789819316641\\
10	0.00952606503598497\\
11	0.0160337954842624\\
12	0.0304713868651603\\
13	0.0404016053317691\\
14	0.0216439644177882\\
15	0.00174941527872807\\
16	0.00540517992051237\\
17	0.0149667983871773\\
18	0.0186787516144942\\
19	0.0215945102349404\\
20	0.00992141837535598\\
21	0.00125235821250735\\
22	0.00813876218385112\\
23	0.0194217149449536\\
24	0.022881381700921\\
25	0.0263648179050753\\
26	0.0157844574525057\\
27	0.0037230553593357\\
28	0.0184512428172532\\
29	0.0249624346793615\\
30	0.0175738513228381\\
31	0.0109817460744106\\
};
\addlegendentry{\revI{$\xi$, \ang{30}}}

\addplot [color=mycolor5, very thick]
  table[row sep=crcr]{%
1	1\\
3	0.224813209758359\\
5	0.12545031522064\\
7	0.0729839609358578\\
9	0.0429165078081049\\
11	0.0257872940317263\\
13	0.0172093876479522\\
15	0.0133689578266889\\
17	0.0117221054055676\\
23	0.00871007643789667\\
25	0.00706488915829695\\
27	0.00615397742165058\\
29	0.00604038063067946\\
31	0.00673856682555751\\
};
\addlegendentry{$\xi_Z$}


\addplot [color=mycolor6, very thick, forget plot]
  table[row sep=crcr]{%
1	1\\
3	0.152697544572298\\
5	0.0821395886957896\\
7	0.0460709945839857\\
9	0.0255213970792878\\
11	0.0146803545368442\\
13	0.0100172528439137\\
15	0.00791110761444823\\
17	0.00605468215734762\\
19	0.00411088455410595\\
21	0.00246574818325593\\
23	0.00171463688064132\\
25	0.00170979969149004\\
27	0.00180914741149408\\
29	0.0015572449625008\\
31	0.00135176126209291\\
}
node[pos=0.63, inner sep=0pt, pin={[pin edge={Stealth-, solid, color=mycolor6}] 230 :$-2.4$}] {};

\addplot [color=mycolor6, very thick]
  table[row sep=crcr]{%
1	1\\
3	0.201363803370536\\
5	0.137954525542748\\
7	0.0966559748095434\\
9	0.0665713152141168\\
11	0.0461948143369305\\
13	0.033937481565379\\
15	0.0271461827040058\\
17	0.0216725351064915\\
19	0.0161945897541895\\
21	0.00975942212949787\\
23	0.00453969477918253\\
25	0.000630777629991399\\
27	0.00287877513435703\\
29	0.00350871897040783\\
31	0.00428180346455564\\
}
node[pos=0.23, inner sep=0pt, pin={[pin edge={Stealth-, solid, color=mycolor6}] above :$-2$}] {};
\addlegendentry{$\xi_S$}

\addplot [black, very thick, dashed, forget plot]
{(0.5*x)^(-2.4} 
node[pos=0.8, inner sep=0pt, pin={[pin edge={Stealth-, solid, color=black}] 240 :$\elDist^{-2.4}$}] {};

\addplot [color=black, very thick, dashed, forget plot]
{(0.5*x)^(-2} 
node[pos=0.36, inner sep=0pt, pin={[pin edge={Stealth-, solid, color=black}] above :$\elDist^{-2}$}] {};

\end{axis}

\end{tikzpicture}%
    \caption{\revII{Normalized} active impedance for the center element in a $N\multX N$ bowtie array steered to boresight (BS) or a \ang{30} tilt in the E-plane. The error is given by the deviation from the asymptotic value. The predictor $\xi_S$ is evaluated for the two convergent behaviors shown in \figref{fig:BTdecay} as indicated by the annotations on the curves.}
    \label{fig:predictor}\label{fig:predictorBT}
\end{figure}

As introduced in \secref{sec:predict} and as \rev{outlined in} \cite{HultinHarald2025IAIi}, passive coupling can give an approximate error on active impedance in the form of $\xi_Z$ from \eqref{eq:predZ}. Applying this to the bowtie array gives the predicted decay shown in \figref{fig:predictor}, that can be compared to the actual decay in the simulated array $\xi$. \revI{The frequency of the bowtie array is chosen such that the array spacing is slightly larger than $\lambda/2$, similar to the numerical results presented in \cite{amitay1972theory}. Furthermore, it gives insight in to how the predictors for spacings other than $\lambda/2$, which is of interest in wideband cases.} The actual array is shown for two different excitations: boresight (\ang{0}) and with the main lobe steered to \ang{30} of boresight in the E-plane. As can been seen, $\xi_Z$ \revIII{predicts $\xi$ well}. While it gives a good prediction, it does require the simulation of two impedances per array size. A predictor without any \revI{additional} simulations would be even more attractive.

\rev{The alternative predictor $\xi_S$ with an expected decay of $\elDist^{-2}$ gives the plot annotated by "$-2$" \revII{in \figref{fig:predictorBT}}}. It can also be seen that the behavior of $\xi_S$ is similar to $\elDist^{-2}$, the decay of expected decay of $S_{cn}$, for large $\elDist$ as expected from \eqref{eq:ZAexp}. Interestingly, $\elDist^{-2}$ is even more similar to the previous predictor $\xi_Z$. \revI{Note that the curve denoted by $\elDist^{-2}$ is actually that function, it is not a curve that is $\propto \elDist^{-2}$.} However, using $\revII{S}$ as a basis of approximation raises an important question: how does $\revII{S}$ actually \revII{decay} in a finite array?

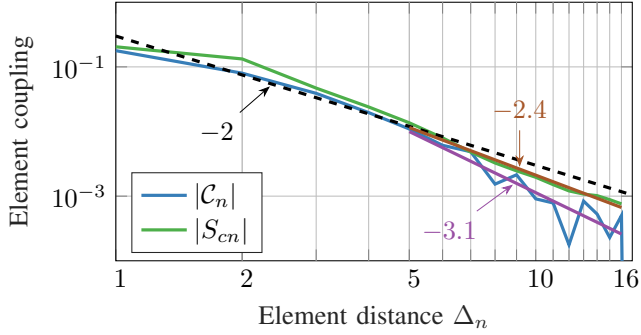
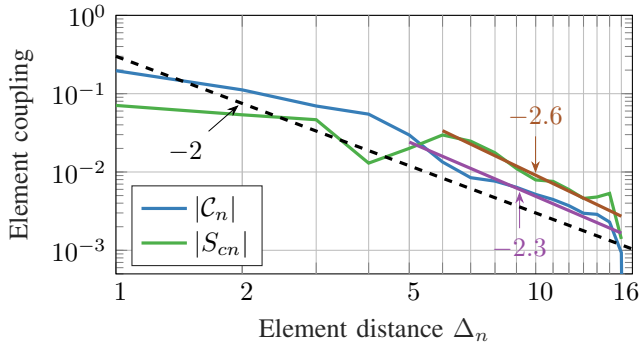
\begin{figure}
    \centering
    \subfloat[\revI{Bowtie array, $31\multX31$ elements at \SI{100}{\mega\hertz}.}]{
%
%
\colorlet{mycolor1}{black}%
\colorlet{mycolor2}{Set1-A}%
\colorlet{mycolor3}{Set1-B}%
\colorlet{mycolor4}{Set1-C}%
\colorlet{mycolor5}{Set1-D}%
\colorlet{mycolor6}{Set1-E}%
\colorlet{mycoloryellow}{Set1-F}%
\colorlet{mycolor7}{Set1-G}%
\begin{tikzpicture}

\begin{axis}[%
width=0.95\columnwidth,
height=5cm,
xmode=log,
xmin=1,
xmax=16.9799931192659,
xminorticks=true,
xtick={1, 2, 5, 10, 16},
log x ticks with fixed point,
extra x ticks={2, 3, 4, 5, 6, 7, 8, 9, 11, 12, 13, 14, 15},
extra x tick label=\empty,
ymode=log,
ymin=1e-04,
ymax=1,
yminorticks=true,
axis background/.style={fill=white},
title style={font=\bfseries},
legend style={at={(0.03,0.03)}, anchor=south west, legend cell align=left, align=left, draw=white!15!black, legend columns=1},
xmajorgrids,
ymajorgrids,
xlabel style={font=\color{white!15!black}},
xlabel={\revII{Element distance} $\elDist$},
ylabel style={font=\color{white!15!black}},
ylabel={\revII{Element coupling}},
domain=1:31,
]
\addplot [color=mycolor3, very thick]
  table[row sep=crcr]{%
1	0.178216564119047\\
2	0.0798652841024414\\
3	0.0389445728924431\\
4	0.0196226746935964\\
5	0.010945053731704\\
6	0.00611964538057628\\
7	0.0048682429907538\\
8	0.00153441159958643\\
9	0.00213889927930313\\
10	0.000907694451853416\\
11	0.000784678334394393\\
12	0.000176747269882974\\
13	0.000837490865234592\\
14	0.000527921541327567\\
15	0.000229319168571856\\
16	0.000499828040806727\\
17	2.88951452049825e-17\\
18	4.48235126059183e-17\\
19	4.03620491081045e-17\\
20	4.99685774487803e-17\\
21	4.9061272478874e-17\\
22	7.01557666725626e-17\\
23	9.7626389499981e-17\\
24	2.74173067991914e-17\\
25	4.35396774376452e-17\\
26	1.51924227779429e-17\\
27	2.96276688432449e-17\\
28	1.9797959815211e-17\\
29	2.62329140248381e-17\\
30	3.01358302890408e-17\\
};
\addlegendentry{\revI{$|\colS{n}|$}}


\addplot [color=mycolor4, very thick]
  table[row sep=crcr]{%
1	0.204521736627259\\
2	0.13265750579818\\
3	0.0469538466640077\\
4	0.0239842823980221\\
5	0.0136307567359058\\
6	0.00796412732653956\\
7	0.00481340518996876\\
8	0.00326011656196293\\
9	0.00246121266863367\\
10	0.00195931076930296\\
11	0.00150698222185234\\
12	0.00120469969682505\\
13	0.00107658528767636\\
14	0.00101756864161487\\
15	0.000883089631481412\\
16	0.00075924767597827\\
};
\addlegendentry{\revI{$|S_{cn}|$}}

\addplot [color=mycolor5, very thick, forget plot]
  table[row sep=crcr]{%
5	0.00997038688425533\\
6	0.00562604314197941\\
7	0.00346810096616809\\
8	0.00228078854753144\\
9	0.00157595469597139\\
10	0.00113222997073213\\
11	0.00083950720927561\\
12	0.000638889416822944\\
13	0.000496964576043293\\
14	0.000393835053845438\\
15	0.000317157741998235\\
16	0.000259004708683435\\
}
node[pos=0.5, inner sep=0pt, pin={[pin edge={Stealth-, solid, color=mycolor5}]225:$-3.1$}] {};


\addplot [color=mycolor1, very thick, dashed, forget plot]
{0.3*x^(-2} node[pos=0.25, inner sep=0pt, pin={[pin edge={Stealth-, solid, color=mycolor1}]225:$-2$}] {};


\addplot [color=mycolor7, very thick, forget plot]
  table[row sep=crcr]{%
5	0.0115231033489503\\
6	0.00737373561060862\\
7	0.00505548417040872\\
8	0.00364556317249418\\
9	0.00273220576971\\
10	0.0021109228749285\\
11	0.00167154825605471\\
12	0.00135079819235731\\
13	0.00111038234973187\\
14	0.00092611658994856\\
15	0.000782164694106304\\
16	0.000667832480519745\\
}
node[pos=0.52, inner sep=0pt, pin={[pin edge={Stealth-, solid, color=mycolor7}] above :$-2.4$}] {};

\end{axis}

\end{tikzpicture}%
        \label{fig:BTdecay}
    }
    \\
    \subfloat[\revI{BoR array, $32\multX32$ elements at \SI{18}{\giga\hertz}.}]{
%
%
\colorlet{mycolor1}{black}%
\colorlet{mycolor2}{Set1-A}%
\colorlet{mycolor3}{Set1-B}%
\colorlet{mycolor4}{Set1-C}%
\colorlet{mycolor5}{Set1-D}%
\colorlet{mycolor6}{Set1-E}%
\colorlet{mycoloryellow}{Set1-F}%
\colorlet{mycolor7}{Set1-G}%
\begin{tikzpicture}

\begin{axis}[%
width=0.95\columnwidth,
height=5cm,
xmode=log,
xmin=1,
xmax=17,
xminorticks=true,
xtick={1, 2, 5, 10, 16},
log x ticks with fixed point,
extra x ticks={2, 3, 4, 5, 6, 7, 8, 9, 11, 12, 13, 14, 15},
extra x tick label=\empty,
ymode=log,
ymin=0.0005,
ymax=1,
yminorticks=true,
axis background/.style={fill=white},
title style={font=\bfseries},
xlabel style={font=\color{white!15!black}},
xlabel={\revII{Element distance} $\elDist$},
ylabel style={font=\color{white!15!black}},
ylabel={\revII{Element coupling}},
xmajorgrids,
ymajorgrids,
legend style={at={(0.03,0.03)}, anchor=south west, legend cell align=left, align=left, draw=white!15!black, legend columns=1},
domain=1:31,
]
\addplot [color=mycolor3, very thick]
  table[row sep=crcr]{%
1	0.196542611167223\\
2	0.111769030445363\\
3	0.0695979361576203\\
4	0.0547619774854701\\
5	0.029640107757321\\
6	0.0134192823902422\\
7	0.00846076699423255\\
8	0.0076465076082943\\
9.00000000000001	0.00637948380422291\\
10	0.00517509671515452\\
11	0.00445582598042722\\
12	0.00371266097146502\\
13	0.00297217571407562\\
14	0.00287905630933769\\
15	0.00229394411554397\\
16	0.000933309600412039\\
16.1029497382883	3.98107170553497e-05\\
};
\addlegendentry{\revI{$|\colS{n}|$}}

\addplot [color=mycolor4, very thick]
  table[row sep=crcr]{%
1	0.070775216889079\\
2	0.0537313377468191\\
3	0.0464963321452264\\
4	0.0129812460297479\\
5	0.02003681328324\\
6	0.0297040074669833\\
7	0.0247557263227198\\
8	0.0175929297532663\\
9	0.0110716011934951\\
10	0.00791679561893706\\
11	0.00758890425141527\\
12	0.00596706292060251\\
13	0.00459492893627346\\
14	0.00479365189155063\\
15	0.00532638252906143\\
16	0.00139716220744737\\
};
\addlegendentry{\revI{$|S_{cn}|$}}

\addplot [color=mycolor5, very thick]
  table[row sep=crcr]{%
5	0.0241020833917674\\
16	0.0016589997263786\\
}
node[pos=0.52, inner sep=0pt, pin={[pin edge={Stealth-, solid, color=mycolor5}] below :$-2.3$}] {};

\addplot [color=mycolor1, very thick, dashed, forget plot]
{0.3*x^(-2} node[pos=0.2, inner sep=0pt, pin={[pin edge={Stealth-, solid, color=mycolor1}]225:$-2$}] {};

\addplot [color=mycolor7, very thick, forget plot]
  table[row sep=crcr]{%
6	0.0336974874519969\\
16	0.00272295497988149\\
}
node[pos=0.52, inner sep=0pt, pin={[pin edge={Stealth-, solid, color=mycolor7}] above :$-2.6$}] {};

\end{axis}
\end{tikzpicture}%
        \label{fig:bordecay}
    }
    \caption{Amplitudes of coupling coefficients as a function of inter-element distance for \revI{two arrays. Here,} $S_{cn}$ is the coupling from the center to each subsequent element along the E-plane in \revI{the array} and \revI{$\colS{n}$} is the predicted coupling from the behavior of the center element using \eqref{eq:amitayPredTheta}. \revI{Lines annotated with $a$ indicate that the line is $\propto \elDist^a$.}}
    \label{fig:decays}
\end{figure}

For a fixed $31\multX 31$ \rev{bowtie} array, $S_{cn}$ decays as $\elDist^{-2.4}$ as shown in \figref{fig:BTdecay}. \revI{Recall that $S$ is the actual passive scattering matrix found by simulating the full array, and $S_{cn}$ is the entry therein corresponding to the coupling from the center element $c$ to another element $n$.} But if this decay is \rev{instead is used in the derivation of \eqref{eq:predS}}, \rev{annotated by "$-2.4$",} this does not \rev{predict the decay} well. Furthermore, if \revI{the entries in} $S$ \revI{along the E-plane are predicted} using \eqref{eq:amitayPredTheta}, denoted by \revI{$\colS{n}$}, the decay is overestimated. The same analysis applied to the BoR array shows a similar result for $S_{cn}$ as seen in \figref{fig:bordecay}, with a slope of $-2.6$ compared to $-2$ of the infinite case. The ARC-based \revI{$\colS{n}$} has a more correct slope for the BoR compared to the bowtie, but has a larger difference in amplitude \revI{as seen in \figref{fig:bordecay}}. The Vivaldi element did not converge to a clear slope within the simulated sizes.

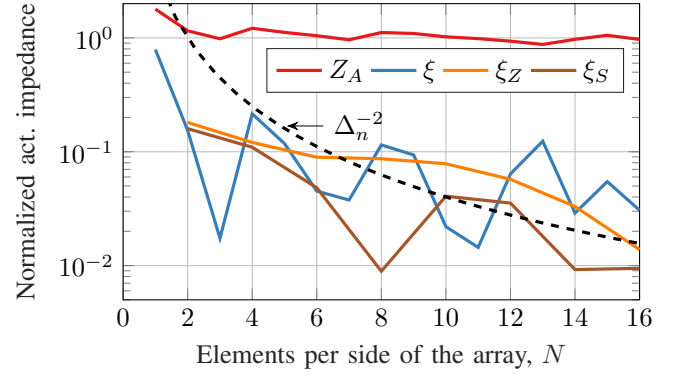
\begin{figure}[t]
    \centering
%
%
\colorlet{mycolor1}{Set1-A}%
\colorlet{mycolor2}{Set1-B}%
\colorlet{mycolor3}{Set1-C}%
\colorlet{mycolor4}{Set1-D}%
\colorlet{mycolor5}{Set1-E}%
\colorlet{mycoloryellow}{Set1-F}%
\colorlet{mycolor6}{Set1-G}%
\begin{tikzpicture}

\begin{axis}[%
width=0.95\columnwidth,
height=5.5cm,
xmin=0,
xmax=16,
xlabel style={font=\color{white!15!black}},
xlabel={\revII{Elements per side of the array,} $N$},
ymin=5*10^-3,
ymax=2,
ymode=log,
ylabel style={font=\color{white!15!black}},
ylabel={Normalized \revII{act.} impedance},
xmajorgrids,
ymajorgrids,
axis background/.style={fill=white},
title style={font=\bfseries},
domain=1:16,
legend style={at={(0.97,0.85)}, anchor=north east, legend cell align=left, align=left, draw=white!15!black, legend columns=4}
]
\addplot [color=mycolor1, very thick]
  table[row sep=crcr]{%
1	1.78976891015311\\
2	1.15385106938674\\
3	0.982601416862558\\
4	1.21497526425343\\
5	1.11786882376996\\
6	1.04511552200574\\
7	0.962348744325009\\
8	1.11477404446478\\
9	1.09367305453986\\
10	1.02198087030866\\
11	0.98557281274854\\
12	0.936168530500463\\
13	0.876084304642287\\
14	0.971085809743315\\
15	1.05465901043835\\
16	0.969399756516413\\
};
\addlegendentry{$Z_A$}

\addplot [color=mycolor2, very thick]
  table[row sep=crcr]{%
1	0.789768910153111\\
2	0.153851069386739\\
3	0.0173985831374424\\
4	0.214975264253429\\
5	0.117868823769957\\
6	0.0451155220057444\\
7	0.0376512556749908\\
8	0.114774044464777\\
9	0.0936730545398596\\
10	0.0219808703086599\\
11	0.0144271872514601\\
12	0.0638314694995366\\
13	0.123915695357713\\
14	0.0289141902566854\\
15	0.0546590104383462\\
16	0.0306002434835868\\
};
\addlegendentry{$\xi$}

\addplot [color=mycolor5, very thick]
  table[row sep=crcr]{%
2	0.180801498252919\\
4	0.120715699943339\\
6	0.0895209542777557\\
8	0.0867692029122757\\
10	0.078408102901605\\
12	0.0573021764074966\\
14	0.0330099078529038\\
16	0.0138161889805701\\
};
\addlegendentry{$\xi_Z$}

\addplot [color=mycolor6, very thick]
  table[row sep=crcr]{%
2	0.15951224334334\\
4	0.1095152504179\\
6	0.0482366549364386\\
8	0.00891105054137498\\
10	0.0406809175233711\\
12	0.0354005395090731\\
14	0.00922474205840729\\
16	0.0094508379168019\\
};
\addlegendentry{$\xi_S$}

\addplot [black, very thick, dashed, forget plot]
{(0.5*x)^(-2} 
node[pos=0.31, inner sep=2pt, pin={[pin edge={Stealth-, solid, color=black}]0:$\elDist^{-2}$}] {};

\end{axis}
\end{tikzpicture}%
    \caption{\revII{Normalized} active impedance for the center element in a $N\multX N$ Vivaldi array steered to boresight (BS). The error is given by the deviation from the convergent value. The predictor $\xi_S$ is evaluated for the expected infinite decay of $\elDist^{-2}$.}
    \label{fig:predictorViv}
\end{figure}

\rev{Given that the Vivaldi array does not reach a convergent behavior within the investigated sizes, a predictor of how it behaves becomes even more attractive. Applying the predictors} to the Vivaldi array gives similar results as for the bowtie array as seen in \figref{fig:predictorViv}. The impedance based predictor $\xi_Z$ follows the deviation well, though drops of a bit towards the edge. On the other hand, $\xi_S$ seems to have two large deviations and \revI{does not} follow the $\elDist^{-2}$ curve as well. Anecdotally, it seems that $\xi_S$ dips around $3\lambda-4\lambda$ away from the array edge, as is the case for the bowtie in \figref{fig:predictorBT} as well. The theoretical decay $\elDist^{-2}$ is also a decent approximation, but does not flatten out in the same manner as the actual error. \revI{As earlier, note that the curve denoted by $\elDist^{-2}$ is actually that function, it is not a curve that is $\propto \elDist^{-2}$.} So similarly to the bowtie, the error in $\revI{Z_a}$ seems to have a slightly shallower slope than the theory, in contrast to $S$ that has a steeper slope than the \rev{infinite array} theory. It seems that $S$ has a slope closer to $-2.5$ rather than $-2$ for \revI{the bowtie and BoR arrays}.

\section{\revIV{Discussion}}
\label{sec:discussion}

\revIV{While the presented predictors give an approximation of the error in active impedance, it is important to consider why such an approximation is valuable. The objective of these predictors is not to replace or make established methods for array antenna modeling obsolete, nor to argue that a certain size of approximation error is acceptable. Rather, these predictors provide a method to estimate the error before committing to a chosen approximation. In this sense, the predictors complement existing rules of thumb by quantifying their validity for a specific configuration. A small predicted error provides additional confidence in the approximation, whereas a large predicted error indicates that the validity of the approximation should be reassessed. The final acceptable error threshold remains application dependent, but the ability to estimate the error beforehand enables a more informed trade-off between computational efficiency and accuracy.}

\revIV{As an example to this application dependence, consider an amplifier that is specified to safely operate at $\Gamma_a < 0.2$. \revV{Further, to facilitate the discussion, the modeling error is assumed to be either \SI{1}{\percent} or \SI{10}{\percent}, representative of the variation} in \figref{fig:normImp}. Suppose an approximation of the array has found that $\Gamma_a=0.167$, corresponding to a safety margin of approximately $\SI{15}{\percent}$. If the modeling error in active impedance is \SI{1}{\percent}, the resulting change in $\Gamma_a$ remains small an the design stays within the specified limits with a worst case of $\Gamma_a=0.172$. In contrast, a \SI{10}{\percent} error would result in a worst case of $\Gamma_a=0.213$, exceeding the specified safety limit. Thereby, the predictors enables for a designer to judge whether the available design margin is sufficient, or whether, e.g., a more accurate model is required.}

\revIV{Established array-modeling methods remain valuable and often perform well in practice. However, their success often relies on engineering experience and intuition regarding when a particular approximation is likely to be accurate. Such intuition can be difficult to formalize and transfer. The proposed predictors complement the established methods by providing a quantitative estimate of the associated modeling error, thereby converting part of this intuition into a measurable quantity that can support modeling decisions.}

\revIV{The methods presented use the electric field integral equation as specified in \cite{akerstedt2024partitioning}, and are therefore formulated to work on structures modeled as Perfect Electric Conductors (PEC). Extension to more general antenna configurations, such as multilayer substrate-based arrays, would require the use of alternative integral equations together with a partitioning scheme that preserves the resulting MBT structure. While such an extension falls outside the scope of the present work, the underlying MLFFT methods and predictor framework are not limited to PEC structures.}

\revIV{For extremely large structures, e.g., for terahertz applications, the electrical size of the problem may become sufficiently large that high-frequency methods or unit cells remain preferable from a computational perspective. Consequently, the proposed methods should be viewed as complementary to, rather than replacements for, existing methods.}

\revV{The examples considered here are regular planar arrays, selected to allow comparison with established array-modeling methods. However, the proposed predictors could also be evaluated for other types of periodic structures. Examples include irregular arrays and conformal arrays, where established rules of thumb are often less readily applicable. In addition, a comprehensive sensitivity analysis over a wide parameter space would provide a better understanding of the robustness and limitations of the proposed predictors. Such a study could identify the parameter ranges for which each predictor is most reliable and further clarify the relationship between predictor accuracy and array configuration. Due to a lack of space, these studies are left for future work.}

\section{Conclusion}
\label{sec:conclusion}
We have demonstrated an improved \revII{full-wave} solver for finite array antennas \revII{that does not rely on} macroscopic assumptions \revII{regarding} element behavior. \revII{A new preconditioning scheme is introduced for MBT matrices arising in finite-array analysis. In addition, we show that the MLFFT of the circulant extension of the MBT matrix commutes with Hermitian conjugation, allowing the Hermitian-conjugate operator to be applied in the transform domain, substantially reducing the memory requirements of the preconditioner.} \revII{The resulting formulation enables the use} of PCG, \revII{which preserves the correspondence between the convergence criterion and the achieved solution accuracy while maintaining low memory requirements.} Furthermore, the \revII{presented} MBT data structure \revII{allows more efficient MLFFT computations and improved overall solver performance}. 

\revII{The} improved solver \revII{has been} used to \rev{investigate the asymptotic \revII{convergence of} the active impedance in finite arrays and to compare the results with established infinite-array theories}. \revII{Several predictors for the convergence of the center-element active impedance and the decay of edge coupling have been examined. The} results show that \revI{the $S$-matrix behavior of} \rev{finite \revI{wideband} bowtie and BoR} arrays \revII{differs from the} \revI{asymptotic} \revII{predictions derived for infinite arrays in \cite{galindo1968asymptotic, amitay1972theory}}, \revII{even for arrays containing up to} 1000 elements. \revII{Thus, the asymptotic regime predicted by infinite-array theory is not reached within the range of array sizes considered here.}

The proposed predictors \revII{provide a practical means of estimating} the error \revII{associated with macroscopic array approximations without requiring simulation of} the full array. \revII{Among the investigated approaches,} \revI{the impedance-based predictor\revV{, $\xi_Z$, demonstrates high predictive accuracy}. The theoretical $\elDist^{-2}$ \revII{decay of} $S$ also \revII{provides accurate predictions of} $Z_a$ without requiring any simulations.} \revII{Together, these predictors provide practical tools for assessing the validity of macroscopic array approximations in large finite-array analyses.}

\appendices
\section{\revIV{Derivation of Hermitian conjugate property}}\label{app:herm}
\revI{To show the \revIV{property of \eqref{eq:multCom}}, first note that} the reciprocity of interactions between mesh cells\revI{\footnote{\revI{The interaction between mesh cell $m_1$ and mesh cell $m_2$ is identical to the interaction between mesh cell $m_2$ and mesh cell $m_1$.}} implies that} the MoM-matrix $Z$ is symmetric, \revII{hence $Z_A$ is also symmetric}. Consider a \revI{generic,} symmetric, 1-level block-Toeplitz matrix $\mli{T}{1}$ consiting of $N\multX N$ blocks, and that is extended to a circulant block-matrix $C(\mli{T}{1})$ as described in \cite{hultin2025solver}:
\begin{equation}
\label{eq:scalarC}
\begin{split}
    C(\mli{T}{1}) 
    =\begin{bmatrix}
    t_{0} & t_1^T & \hdots & t_{N-1}^T & t_{N-1} & \hdots & t_{1} \\
    t_{1} & \ddots & \ddots & \ddots & \ddots & \ddots & \ddots \\
    \vdots & \ddots & \ddots & \ddots & \ddots & \ddots & \ddots \\
    t_{N-1} & \ddots & \ddots & \ddots & \ddots & \ddots & \ddots \\
    t_{N-1}^T & \ddots & \ddots & \ddots & \ddots & \ddots & \ddots  \\
    \vdots & \ddots & \ddots & \ddots & \ddots & \ddots & \ddots \\
    t_{1}^T & \ddots & \ddots & \ddots & \ddots & \ddots & \ddots  \\
    \end{bmatrix},
\end{split}
\end{equation}
where $t_n$ is a \revI{$l\multX l$} matrix. \revI{As in \eqref{eq:multL},} \rev{$C_1(\mli{T}{1})$ is the first column in \eqref{eq:scalarC}}\revII{.} \revI{The blocks} \rev{along the first column \revI{$C_1$}} \revI{will be refereed to as $c_{n}$}. \revI{The symmetry of $\mli{T}{1}$ gives}
\begin{equation}
\label{eq:CHerm}
\begin{split}
    C((\mli{T}{1})^H)  = C(\mli{T}{1})^*.
\end{split}
\end{equation}
This also implies that
\begin{equation}
\label{eq:FHerm}
\begin{split}
    \mli{\mathcal{F}}{1}\left[C((\mli{T}{1})^H)\right]  = \mli{\mathcal{F}}{1}\left[C(\mli{T}{1})^*\right].
\end{split}
\end{equation}

The 1-level block-Fourier transform block diagonalizes $C(\mli{T}{1})$ as \revII{\cite{hultin2025solver, bleszynski2004block}}
\begin{equation}
    \mathcal{F}^{[1]}(C(\mli{T}{1}))\left(\mathcal{F}^{[1]}\right)^{-1} = \text{diag}_K (\mathcal{F}^{[1]}(C_1(\mli{T}{1}))).
\end{equation}
Due to the extension to a circulant matrix, $K=2N-1$  \cite{hultin2025solver}.

For the left-hand side of \eqref{eq:FHerm} the $k$:th block becomes
\begin{equation}
\label{eq:FCC}
    \begin{split}
    [\mathcal{F}^{[1]}(C_1((\mli{T}{1})^H))]_k=\\
    [\mathcal{F}^{[1]}(C_1((\mli{T}{1}))^*)]_k = \sum^{2N-2}_{n=0} \{ \revI{c_{n}^*}\}\e ^{-\ju2\pi\frac{k}{2N-1}n}= \\ t_0^*+t_1^*\e ^{-\ju2\pi\frac{k}{2N-1}}+\dots+t_{N-1}^*\e ^{-\ju2\pi\frac{k(N-1)}{2N-1}}\\+t_{N-1}^H\e ^{-\ju2\pi\frac{kN}{2N-1}}+\dots+t_1^H\e ^{-\ju2\pi\frac{k(2N-2)}{2N-1}}, \\ k=1,\dots,K,
    \end{split}
\end{equation}
The right-hand side of \eqref{eq:FHerm} is
\begin{equation}
\label{eq:FHC}
    \begin{split}
    [\mathcal{F}^{[1]}(C_1(\mli{T}{1}))]_k^H= \sum^{2N-2}_{n=0} \{ \revI{c_{n}^H}\}\left(\e ^{-\ju2\pi\frac{k}{2N-1}n}\right)^*= \\ t_0^*+t_1^H\left(\e ^{-\ju2\pi\frac{k}{2N-1}}\right)^*+\dots+t_{N-1}^H\left(\e ^{-\ju2\pi\frac{k(N-1)}{2N-1}}\right)^*\\+t_{N-1}^*\left(\e ^{-\ju2\pi\frac{kN}{2N-1}}\right)^*+\dots+t_1^*\left(\e ^{-\ju2\pi\frac{k(2N-2)}{2N-1}}\right)^*, \\ k=1,\dots,K.
    \end{split}
\end{equation}
Using that the Fourier coefficients are equispaced around the unit circle, i.e. $\e ^{-\ju2\pi\frac{kn}{2N-1}}=\left(\e ^{-\ju2\pi\frac{k(2N-1-n)}{2N-1}}\right)^*$, \eqref{eq:FHC} can be rewritten as
\begin{equation}
\label{eq:FHC2}
    \begin{split}
    [\mathcal{F}^{[1]}(C_1(\mli{T}{1}))]_k^H=\\ t_0^*+t_1^H\e ^{-\ju2\pi\frac{k(2N-2)}{2N-1}}+\dots+t_{N-1}^H\e ^{-\ju2\pi\frac{kN}{2N-1}}\\+t_{N-1}^*\e ^{-\ju2\pi\frac{k(N-1)}{2N-1}}+\dots+t_1^*\e ^{-\ju2\pi\frac{k(2N-2)}{2N-1}}= \\
    [\mathcal{F}^{[1]}(C_1((\mli{T}{1})^H))]_k, \: k=1,\dots,K.
    \end{split}
\end{equation}
Thus \eqref{eq:FCC} is identical to \eqref{eq:FHC2}, and it is demonstrated that
\begin{equation}
    \mathcal{F}^{[1]}(C(T_1)^H)\left(\mathcal{F}^{[1]}\right)^{-1} = \text{diag}_K (\mathcal{F}^{[1]}(C_1(T_1)))^H.
\end{equation}

\section{\revIV{Scaling of $S$ and \revV{$Z_p$} entries}}
\label{app:scaling}
\revV{Given that $Z_p$ in \eqref{eq:S2Z} is well defined, $\id-S$ must be invertible}. Therefore, \revV{$(\id-S)^{-1}$} can be described by a Neumann series
\begin{equation}
\label{eq:neumann}
    (\id-S)^{-1} = \sum_{k=0}^\infty S^k.
\end{equation}
While this holds for the whole $S$-matrix, it may converge slowly in arrays with high coupling. For the sake of analysis, divide $S$ into a matrix with elements of large coupling $S_L$ and weak coupling $S_W$, usually corresponding to interactions that are close and far apart respectively, such that $S=S_L+S_W$. Then the inverse can be expanded as
\begin{align}
\label{eq:Sexp}
    \begin{aligned}
        (\id-S)^{-1} &= \inv{\left[\id-S_L-S_W\right]}\\&=\inv{\left[(\id-S_W\inv{(\id-S_L)})(\id-S_L)\right]}\\&=\inv{(\id-S_L)}\inv{\left((\id-S_W\inv{(\id-S_L)})\right)}.
    \end{aligned}
\end{align}

Inserting \eqref{eq:Sexp} into \eqref{eq:S2Z} together with \eqref{eq:neumann} gives
\begin{align}
\begin{aligned}
    \revI{Z_p}=Z_0[ &(\id+S_L+S_W)\inv{(\id-S_L)}\\ &\left(\id+S_W\inv{(\id-S_L)}+\mathcal{O}(S_W^2)\right) ]\\
    = Z_0[&\revII{\zeta}_L+(\id+\revII{\zeta}_L)S_W\inv{(\id-S_L)})+\mathcal{O}(S_W^2)],
\end{aligned}
\end{align}
where $\revII{\zeta}_L=(\id+S_L)(\id-S_L)^{-1}$. \rev{Thus $\|\revI{Z_p}-Z_0\revII{\zeta_L}\|_2\leq \revI{G_1}\|S_W\|_2$, where \revI{$G_1$} is \revI{some} constant. Hence the growth rate of added elements in $\revI{Z_p}$ should} \revI{be similar to the growth rate of added elements in $S$}\rev{, \revV{as the strong coupling introduced by edge elements should have a small effect on couplings between center and edge,} assuming that $\|(\id-S_L)u\|_2\geq \revI{G_2}\|u\|_2$.} The active impedance $\Za{c}$ of the center element $c$ for a boresight excitation, excluding higher order terms, becomes
\begin{multline}
    \label{eq:ZAexp}
    \hat{e}_c^T\revI{Z_a} = \hat{e}_c^T\text{diag}(I_{\revV{p}})^{-1} \revII{Z_p} I_{\revV{p}}\\ \revI{\approx} Z_0\revV{I_{p,c}^{-1}}[\hat{e}_c^T\revII{\zeta}_L+\hat{e}_c^T(\id+\revII{\zeta}_L)S_W\inv{(\id-S_L)})]I_{\revV{p}},\\\rev{\revV{\implies} \|\hat{e}_c^T\revI{Z_a}-\hat{e}_c^TZ_0\revV{I_{p,c}^{-1}}\revII{\zeta}_LI_{\revV{p}}\|_{\revII{2}}\leq \revI{G_3}\|\revV{\hat{e}_c^TI_{p,c}^{-1}}S_WI_{\revV{p}}\|_2},
\end{multline}
\revI{where $G_3$ is some constant.} \revV{The multiplication $\hat{e}_c^TS_W$ extracts the column corresponding to the center element. Adding a new element to a large array is expected to only add weak entries to $S$, and hence $S_W$, while the corresponding column in $S_L$ remains unaffected.} \revV{Given that \eqref{eq:S2Z} is well defined, $\|S\|_{\revII2}<1\implies\|S_L\|_{\revII2}<1$.} \revI{Therefore $\id-S_L$ is invertible, i.e. only contributes a constant norm.}  As the added elements to \eqref{eq:ZAexp} scale linearly with the added elements to $S_W$, the change in the active impedance is also expected to scale similarly as the added elements to \rev{$S_W$} .

\bibliographystyle{IEEEtran}
\bibliography{IEEEabrv, sources.bib}

\end{document}